\documentclass[3p,twocolumn]{elsarticle}

\usepackage{bm}
\usepackage{caption}
\usepackage{fullpage}
\usepackage{amsmath}
\usepackage{amssymb}
\usepackage{graphicx}
\usepackage{indentfirst}
\usepackage{setspace}
\usepackage{color}
\usepackage{hyperref}
\usepackage{cleveref}
\usepackage{verbatim}
\usepackage{subcaption}
\graphicspath {{densityVis/}}
\graphicspath {{plots/}}
\graphicspath {{filterVsDNS/}}
\graphicspath {{div/}}
\graphicspath {{vorticalPlots/}}
\begin{document}
\begin{frontmatter}
\title{Revisiting the Late-Time Growth of Single-mode Rayleigh-Taylor Instability and the Role of Vorticity}

\author[URME]{Xin Bian\corref{mycorrespondingauthor}}
\cortext[mycorrespondingauthor]{Corresponding author}
\ead{xin.bian@rochester.edu}
\author[URME,LLE]{Hussein Aluie}
\author[URME,LLE]{Dongxiao Zhao}
\author[URME,LLE]{Huasen Zhang}
\author[LANL]{Daniel Livescu}
\address[URME]{Department of Mechanical Engineering, University of Rochester, NY 14627, USA}
\address[LLE]{Laboratory for Laser Energetics, University of Rochester, NY 14627, USA}
\address[LANL]{Los Alamos National Laboratory, Los Alamos, NM 87545, USA}

\begin{abstract}
Growth of the single-fluid single-mode Rayleigh-Taylor instability (RTI) is revisited in 2D and 3D using fully compressible high-resolution simulations. We conduct a systematic analysis of the effects of perturbation Reynolds number ($Re_p$) and Atwood number ($A$) on RTI's late-time growth. Contrary to the common belief that single-mode RTI reaches a terminal bubble velocity, we show that the bubble re-accelerates when $Re_p$ is sufficiently large, consistent with [Ramaparabhu et al. 2006, Wei and Livescu 2012]. However, unlike in [Ramaparabhu et al. 2006], we find that for a sufficiently high $Re_p$, the bubble's late-time acceleration is persistent and does not vanish. Analysis of vorticity dynamics shows a clear correlation between vortices inside the bubble and re-acceleration. Due to symmetry around the bubble and spike (vertical) axes, the self-propagation velocity of vortices points in the vertical direction. If viscosity is sufficiently small, the vortices persist long enough to enter the bubble tip and accelerate the bubble [Wei and Livescu 2012]. A similar effect has also been observed in ablative RTI [Betti and Sanz 2006]. As the spike growth increases relative to that of the bubble at higher $A$, vorticity production shifts downward, away from the centerline and toward the spike tip. We modify the Betti-Sanz model for bubble velocity by introducing a vorticity efficiency factor $\eta=0.45$ to accurately account for re-acceleration caused by vorticity in the bubble tip. It had been previously suggested that vorticity generation and the associated bubble re-acceleration are suppressed at high $A$. However, we present evidence that if the large $Re_p$ limit is taken first, bubble re-acceleration is still possible. Our results also show that re-acceleration is much easier to occur in 3D than 2D, requiring smaller $Re_p$ thresholds.
\end{abstract}


\end{frontmatter}

\section{INTRODUCTION}
The Rayleigh-Taylor instability (RTI) appears at a perturbed interface when a light fluid is accelerated against a heavy fluid \cite{PLMS:PLMS0170,taylor1950instability}. RTI is important in many engineering applications such as inertial confinement fusion (ICF) \cite{betti2016inertial} where it can significantly degrade a target's performance \cite{Woo:2018fj,Woo:2018fm}. It also plays an important role in the evolution of astrophysical systems, such as supernova explosions \cite{arnett1989supernova} and gaseous hydrogen clouds \cite{Remingtonetal06}. There has been significant theoretical, experimental, and numerical advances toward understanding the fundamental physics of this problem \cite{sharp1984overview,dimonte2004comparative,Livescuetal10,Livescu2013,boffetta2017incompressible}. Refs. \cite{ZHOU20171,zhou2017rayleigh} offer an extensive recent review on the topic. 

At early times, when the instability amplitude is sufficiently small, the flow is well-described by linear analysis \cite{PLMS:PLMS0170,taylor1950instability}. For the simple incompressible, inviscible, immiscible case in a domain much larger than the perturbation wavelength, the perturbation at the interface grows exponentially with a rate $\gamma=\sqrt{Akg}$, where $g$ is acceleration, $k=2\pi/\lambda$ is perturbation wavenumber ($\lambda$ is wavelength), and $A=(\rho_h-\rho_l)/(\rho_h+\rho_l)$ is the Atwood number ($\rho_h$ and $\rho_l$ are the density of the heavy and light fluids, respectively).  Numerous studies have addressed the the roles of viscosity, mass diffusivity, finite domains, compressibility, background stratification, surface tension, etc., e.g., \cite{DHH62,livescu2004compressibility,LCG07,Rollin_2011,gerashchenko2016}. At later times, when the instability amplitude exceeds $\gtrsim 0.1 \lambda$, nonlinear interactions become important and the flow develops interpenetrating bubbles (due to the light fluid rising) and spikes (due to the heavy fluid sinking).

The nonlinear stage in single-mode RTI has been studied by many analytic models
\cite{Davies375,layzer1955instability,Youngs89,youngs1991three,Alonetal95,zhang1998analytical,oron2001dimensionality,Chengetal02,goncharov2002analytical,Kokkinakisetal19}. An important early contribution was by Layzer \cite{layzer1955instability}, whose model relied on assuming a potential flow away from the fluid-vacuum interface in RTI flows with $A=1$. Goncharov \cite{goncharov2002analytical} later generalized Layzer's theory to arbitrary Atwood numbers but still relying on the  potential flow assumption. The model predicts a terminal bubble velocity of
\begin{equation}
U_{B} = \sqrt{\frac{2\,A\,g}{(1+ A)\,C\,k}},
\end{equation}
where $C=3$ in 2D and $C=1$ in 3D. The model also yields a terminal spike velocity $U_{S} = \sqrt{\frac{2Ag}{(1- A)C\,k}}$, although its validity breaks down due to vorticity generation near the spike tip, violating the potential flow assumption \cite{goncharov2002analytical}. 

The growth of the bubble and the extent of its penetration into the heavy fluid has been the focus of many studies due to the critical role it plays in ICF implosions by mixing ablator material into the fuel. This can have severe adverse effects on the ICF target performance \cite{Edwards:2013hm,Sangster:2013es,Igumenshchev:2013cy,Clark:2016jj,betti2016inertial}. Recent experiments and numerical simulations \cite{betti2006bubble,ramaprabhu2006limits,wilkinson2007experimental,ramaprabhu2012late,wei2012late,yan2016three,Zhangetal18pre,zhang2018self,Xinetal19} have shown that even the prediction of a terminal bubble velocity breaks down at late times. In ablative RTI, it was found that the bubble is accelerated to velocities above the potential flow prediction after a quasi constant-velocity phase \cite{betti2006bubble}.
This re-acceleration was attributed to vortices generated near the spike and advected by the ablative flow toward the bubble tip. The vortices then exert a centrifugal force against the bubble tip causing its re-acceleration. 

Rampamprabhu et al. \cite{ramaprabhu2006limits,ramaprabhu2012late} used Implicit Large Eddy Simulations (ILES) to show that a similar phenomenon occurs in classical RTI at low Atwood numbers, where secondary Kelvin-Helmholtz instabilities (KHI) are responsible for vorticity generation which leads to bubble re-acceleration. However, Refs. \cite{ramaprabhu2006limits,ramaprabhu2012late} observed that the bubble velocity increase above the ``terminal'' value is only transient and that the bubble experiences an eventual deceleration which slows down the bubble back to its terminal velocity at later times. Moreover, Rampamprabhu et al. \cite{ramaprabhu2006limits,ramaprabhu2012late} observed that bubble re-acceleration is completely suppressed for high density ratios with $A\ge 0.6$. It is worth noting the simulations in \cite{ramaprabhu2006limits,ramaprabhu2012late}, while being ground breaking at the time, were at a relatively low cross-sectional grid-resolution of $128^2$ by today's standards. This is especially relevant since the simulations, being ILES, had significant dissipation from the numerical discretization. Indeed, visualizations in \cite{ramaprabhu2006limits,ramaprabhu2012late} indicate that their RTI flows do not preserve the symmetry across the (vertical) bubble axis which indicates a violation of momentum conservation. In the absence of a horizontal force (gravitational acceleration is vertical), momentum conservation necessitates that the center of mass of the entire domain remain along the same vertical line. A break in the left-right symmetry for single-mode RTI implies a horizontal shift in the center of mass.

Direct numerical simulations of 2D single-mode RTI at $A=0.04$ in the incompressible variable density limit \cite{wei2012late} showed that when symmetry around the bubble (vertical) axis is maintained, the self-propagation velocity of vortices points in the vertical direction. This enhances the background vertical advection of vortices and leads to their efficient propagation into the bubble tip, resulting in re-acceleration.
When viscosity is sufficiently small, the vortices last long enough to reach the bubble tip, where the induced vortical velocity brings in less mixed fluid from the interior of the layer and accelerates the bubble. The role of viscosity was quantified using the perturbation Reynolds number $Re_p\equiv {\lambda\sqrt{\frac{A}{1+A}g\lambda}/\nu}$, where $\nu$ is kinematic viscosity and the Atwood number dependency follows Goncharov's result \cite{goncharov2002analytical}. Ref. \cite{wei2012late} showed that bubbles experience different growth stages at low and high $Re_p$. Above a threshold $Re_p$ value, RTI undergoes re-acceleration followed by what they termed a ``chaotic development'' stage where the bubble front's mean acceleration is constant, corresponding to quadratic growth in mixing layer width. In the work of Wei and Livescu \cite{wei2012late}, the chaotic development stage at high $Re_p$ refers to the late-time quadratic growth stage ($h_B\propto gt^2$), which is different to the ``chaotic mixing'' stage in Ref. \cite{ramaprabhu2012late} used to describe the break of symmetry across the (vertical) bubble axis at late times.

However, at high Atwood numbers, as the growth of bubble and spike becomes asymmetric, the largest vorticity production moves downward, away from the (initial interface) center-line and toward the spike tip. Consequently, the vortices need to travel larger distances to reach the bubble tip. Results from Ref. \cite{wei2012late} suggest that higher Atwood numbers require smaller viscosities and that the numerical fidelity of simulations maintain symmetry around the bubble (vertical) axis to observe bubble re-acceleration.
Nevertheless, no DNS study has been performed to date to test this hypothesis.

Despite the important work in the aforementioned studies, several issues remain in the late-time behavior of single-mode RTI which motivate our paper: 

(1) Does the bubble decelerate back to a constant velocity growth after a transient re-acceleration as suggested in \cite{ramaprabhu2012late} and, if so, is this constant velocity similar to the ``terminal velocity''? Moreover, are subsequent re-accelerations possible after a failed re-acceleration? Since the RTI flows in \cite{ramaprabhu2012late} were under-resolved and lost symmetry at late-times, the persistence of bubble re-acceleration deserves revisiting.

(2) Is there a threshold Atwood number above which re-acceleration is completely suppressed as suggested in \cite{ramaprabhu2012late}? Wei and Livescu \cite{wei2012late} studied the effects of varying $Re_p$ at a fixed $A=0.04$ using incompressible 2D DNS, while in ICF the $A\approx 1$ \cite{craxton2015direct}. Other studies investigating this issue \cite{reckinger2016comprehensive,wieland2018} were also at a fixed low Atwood number. 

(3) Are there fundamental differences in RTI bubble growth between 2D and 3D? 
Previous studies of late-time bubble growth were either restricted to 2D \cite{wei2012late,reckinger2016comprehensive,wieland2018} or included 3D but were numerically under-resolved \cite{ramaprabhu2006limits,ramaprabhu2012late}.
The absence of vortex stretching in 2D RTI makes the underlying vortex dynamics significantly different from 3D RTI.  

(4) Will the dynamics of re-acceleration and chaotic development shown in \cite{wei2012late} using incompressible fluid simulations be similar when using the fully compressible dynamics? Compressible effects such as background stratification, finite acoustic speed, non-zero velocity divergence, and the fluid's equation of state can have non-trivial effects on the RTI development \cite{Livescu:2004bw,wieland2017,gauthier2017compressible,wieland2018}. Ref. \cite{olson2007rayleigh} found that rising bubbles compress the heavy fluid and generate shocklets which merge with each other into a normal shock. A compressible code using Parallel Adaptive Wavelet Collocation Method (PAWCM) was developed \cite{reckinger2016comprehensive} and showed that background stratification tends to suppress both bubble and spike growth at $A=0.1$ using 2D DNS. More recently, the effects of isothermal background stratification on 2D compressible RTI at $A=0.04$ were studied by Wieland et. al \cite{wieland2018} using PAWCM, where they showed the perturbation baroclinic torque is the main driver for RTI growth. In their 2D simulations, re-acceleration was observed for weak but not strong stratification. As noted in \cite{wieland2018}, it is still not clear whether the chaotic development stage will appear or not in fully compressible simulations due to the small computational domain in their study. The instability suppression does not appear when the initial configuration has constant density on each side of the interface \cite{wieland2017}.

In this paper, we perform high-resolution fully compressible simulations that maintain symmetry until very late times. The late-time behavior of bubbles and spikes is studied systemically at both low and high Atwood numbers for different $Re_p$. To avoid the complications due to instability suppression in the presence of background stratification, the initial configuration is out of thermal equilibrium, with constant density on each side of the interface. A comparison between 2D and 3D RTI is also conducted. The results show that the re-acceleration and chaotic development stages appear if $Re_p$ is above a threshold which depends on $A$. Higher $A$ have higher $Re_p$ thresholds. This is consistent with the incompressible 2D DNS results of \cite{wei2012late} which investigated $Re_p$ at a fixed $A$. Furthermore, at sufficiently high $Re_p$, the increase in bubble velocity above its ``terminal'' value is persistent and does not decrease. Our analysis indicates that bubble re-acceleration above its ``terminal velocity'' will occur even in the $A\to1$ limit if the $Re_p\to\infty$ limit is taken first.
A comparison between 2D and 3D RTI shows that RTI grows much faster and requires a smaller  $Re_p$ threshold to re-accelerate in 3D than in 2D. 

The paper is organized as follows. A brief description of the governing equations, initialization, and numerical schemes is presented in Section \ref{sec:numerical}. Section \ref{sec:result} discusses the main results, the effects of $Re_p$, $A$, and 3D. We conclude with Section \ref{sec:conclusions}. The effects of filtering in our numerical scheme are compared to DNS results in \ref{app:filterDNScompare}.

\section{NUMERICAL METHODOLOGY}
\label{sec:numerical}
\subsection{Governing Equations}

The numerical simulations are conducted with the DiNuSUR code which has been used in many previous studies (e.g. \cite{zhao2018inviscid,zhang2018self,Lees:2019hn,Bian:2019ip}). We use sixth-order compact finite differences \cite{lele1992compact} in the vertical direction and pseudo-spectral method in the horizontal direction, similar to previous RTI DNS in the variable density limit \cite{cook2001transition,livescu2011direct}. Time integration uses fourth-order Runge-Kutta. We solve the single-fluid compressible Navier-Stokes equations over a Cartesian grid, including the continuity equation (\ref{eq:continuity}), momentum transport (\ref{eq:momentum}), and  total energy transport (\ref{eq:total}):  
\begin{eqnarray}\label{eq:navier-stokes}
& \partial_t\rho+\partial_j(\rho u_j)=0 \label{eq:continuity},\\
&\partial_t(\rho u_i)+\partial_j(\rho u_i u_j)=-\partial_i P\nonumber\\
&+ \partial_j\sigma_{ij}-\rho g\delta_{iz} \label{eq:momentum},\\
& \partial_t (\rho E) + \partial_j (\rho E u_j)=-\partial_j (Pu_j)\nonumber\\
&+ \partial_j (u_i \sigma_{ij}) -\partial_j q_j -\rho u_i g\delta_{iz} \label{eq:total},
\end{eqnarray}
where $\rho$ is density, $u$ is velocity, $P$ is pressure, $\mathbf{g}$ is the gravitational acceleration along the vertical direction $z$, $E=|{\bf u}|^2/2 + e$ is the specific total energy per unit mass, with $e$ the specific internal energy. The viscous stress $\sigma_{ij}$ is defined as
\begin{equation}
\sigma_{ij} = 2\mu (S_{ij}-\frac{1}{3}S_{kk} \delta_{ij}),
\end{equation}
where $S_{ij} = (\partial_ju_i+\partial_iu_j) /2$ is the symmetric strain tensor. The heat flux $q_{j}$ is defined as, $q_j=-\kappa \partial_j T,$ where $T$ is temperature and $\kappa$ is thermal conductivity. The ideal gas law, $P=\rho R T$, $e=R/(\Gamma-1) T$ is used, with the gas constant, $R$, and the ratio of specific heats, $\Gamma$. $R$, $\Gamma$, $\mu$ and $\kappa$ are constants in space and time.

\subsection{Comparison to Previous Studies}

Compared to the previous studies of late-time behavior in single-mode RTI, \cite{wei2012late} (two-fluid incompressible) and \cite{reckinger2016comprehensive} (two-fluid compressible), the compressible single-fluid model is used in this work, leading to differences in background stratification,  acoustic wave generation, and baroclinic vorticity production.

\subsubsection{Background Stratification}
To avoid the instability suppression due to background stratification, the initial density field, $\rho_0$, is uniform on each side of the interface. The hydrostatic equilibrium then requires that, away from the interface, the initial (background) pressure varies as,

\begin{equation}
P_0\sim \rho_0 g z,
\end{equation}
where $z$ is the vertical position. For the single fluid case, using the ideal gas equation of state yields that the background temperature gradient is constant and equal on both sides of the interface. Thus, the initial conditions represent a particular case of the analysis of Ref. \cite{gerashchenko2016}, with $d T_0/dz= g/R$. Away from the interface, when thermal conductivity coefficient is constant, a constant temperature gradient implies that the heat conduction term vanishes in the energy equation, so that the initial conditions are also in thermal equilibrium.

\subsubsection{Initial acoustic effects}

At the interface, as density changes between the heavy and light regions (see below), the temperature gradient can no longer be constant. The energy equation then has non-zero time derivative at initial time,
\begin{equation}
\partial_t (\rho e) = \partial_j (\kappa \partial_j T_0),
\end{equation}
which results in the generation of acoustic waves at initial time.

The generation mechanism of acoustic waves in two-fluid miscible RTI \cite{reckinger2016comprehensive,wieland2018} is different. In the latter case, it is the enthalpy diffusion term in the energy equation leading to non-zero time derivative and the generation of acoustic waves, 
\begin{equation}
\partial_t (\rho e) = -\partial_j (c_{pl}Ts_{jl}),
\end{equation}
where $c_{pl}$ is specific heats at constant pressure of $l$ fluid, $s_{jl}=\rho D \partial_j Y_l$  ($D$ is the mass diffusion coefficient, $Y_l$ is mass fraction of $l$ fluid). By adding an initial dilatational velocity consistent with the incompressible variable density limit \cite{sandoval1995dynamics,Livescu2013}, 
\begin{equation}
\nabla \cdot \bm u = - D \nabla^2 \ln \rho,
\end{equation}
Ref. \cite{reckinger2016comprehensive} was able to minimize the generation of initial acoustic waves. In a similar manner, it is possible to initialize with a dilatational velocity that is consistent with the heat conduction term, i.e. $ \nabla \cdot \bm u = \kappa \nabla^2 T$, to minimize initial acoustic wave generation. Investigating the role of such initializations on RTI growth has not been explored here and is beyond our focus. A detailed survey of similarities and differences between flows with density variations due to thermal and compositional changes is presented in Ref. \cite{livescu2020}.

\subsubsection{Baroclinic Term}
The vorticity equation in this paper (compressible single-fluid) is as follows,
\begin{eqnarray}  \label{eq:vortex}
&\partial_t\bm{\omega}+(\bm{u}\cdot \nabla)\bm{\omega}=(\bm{\omega}\cdot\nabla)\bm{u}-\bm{\omega}(\nabla\cdot \bm{u})\nonumber\\
&+\frac{1}{\rho^2}\nabla\rho\times\nabla P+\nabla\times(\frac{\nabla\cdot\sigma}{\rho}).
\end{eqnarray}
The baroclinic term, $\nabla \rho \times \nabla P$, is essential for the instability growth. While the form of the term is the same in the two-fluid compressible and incompressible, as well as single-fluid compressible RTI cases, the dynamics are potentially different. Thus, for the compressible cases, the pressure density relation is mediated by the equation of state. For the two-fluid configuration the mixture gas constant varies in both space and time so that mixing also plays a role on baroclinic generation. 
There is no ideal gas law relation between pressure and density in two-fluid incompressible RTI \cite{wei2012late}. The additional constraint is from the non-zero velocity divergence shown above.

\subsection{Dimensionless Parameters}
Wei and Livescu \cite{wei2012late} have shown that the development stages of incompressible single-mode RTI are strongly affected by a perturbation Reynolds number defined as $\lambda\sqrt{\frac{A}{1+A}g\lambda}/\nu$, where $\lambda$ is the initial wavelength. In this paper, the late-time behavior is studied with Prandtl number $Pr=\nu/\alpha=1$, where $\alpha=\kappa/(c_p\rho)$ is thermal diffusivity. $Re_p$ here is defined using the interfacial density $\rho_I=(\rho_h+\rho_l)/2$ as,
\begin{equation}
Re_p \equiv \frac{\lambda\sqrt{\frac{A}{1+A}g\lambda}}{\mu/\rho_I}.
\end{equation}
In the Reynolds number definition, $\lambda$ is the perturbation wavelength. $\sqrt{\frac{A}{1+A}g\lambda}$ is a characteristic velocity proportional to the terminal velocity from potential flow theory \cite{goncharov2002analytical}. To quantify the simulation grid resolution, we use the grid Grashoff number \cite{cabot2006reynolds} $Gr_{\Delta}\equiv 2Ag\Delta^3/\nu^2$, where $\Delta$ is the mesh size. $Gr_{\Delta} \le 1$ has been used to denote well-resolved simulations \cite{cabot2006reynolds,wei2012late}. The $Gr_{\Delta}$ in the simulations is shown in Tables \ref{tbl:2dSimpara} and \ref{tbl:3dSimpara} in \ref{app:simpara}. To compare the RTI evolution under different parameters, we use the non-dimensional time $\tau=t\sqrt{Ag/\lambda}$, and dimensionless bubble/spike front velocity, also called Froude number, $Fr_{B/S} \equiv U_{B/S}/\sqrt{\frac{A}{1+A}g \lambda}$,  where $U_{B/S}$ is the dimensional bubble/spike velocity. Note that our dimensionless time, $\tau$, is smaller than the definition used in Ramaprabhu et al. \cite{ramaprabhu2006limits,ramaprabhu2012late} by a factor $\sqrt{2\pi}$. For example, $\tau=6$ in this paper corresponds to dimensionless time $\approx15$ in \cite{ramaprabhu2006limits,ramaprabhu2012late}.

\begin{figure}
\centering
\begin{subfigure}{0.41\textwidth}
\includegraphics[height=1.9 in]{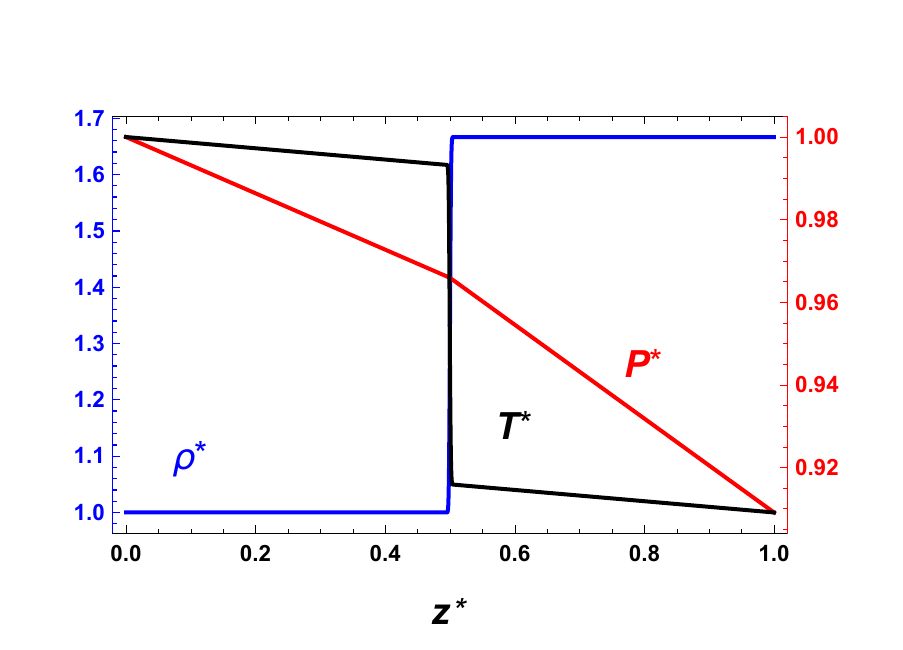}
\end{subfigure}
\begin{subfigure}{0.05\textwidth}
\includegraphics[height=1.9 in]{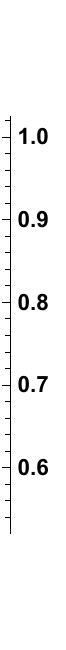}
\end{subfigure}
\renewcommand{\figurename}{FIG.}
\caption{Initial conditions for dimensionless density $\rho^* = \rho/\rho|_{z=0}$ (Blue), $T^* = T/T|_{z=0}$ (black), and pressure $P^*=P/P|_{z=0}$ (Red) along $z^*=z/L_z$ in a $A=0.25$ simulation. $\rho|_{z=0}$, $T|_{z=0}$ and $P|_{z=0}$ are the initial density, temperature, and pressure at the bottom ($z=0$), respectively.}

\label{fig:IC}
\end{figure}

A computational domain with aspect ratio 8 is used in both 2D and 3D simulations. The physical size of the domain is $L_x\times L_z=0.4\times 3.2$ in 2D and $L_x\times L_y\times L_z=0.4\times 0.4\times 3.2$ in 3D. The initial perturbation is at the middle of the computational box, $z_0 = 0.5L_z$, where $L_z$ is the height of the computational box. For high $A$ simulations ($A\ge0.6$), the interface is shifted up to $z_0 = 0.75L_z$ to allow for longer temporal evolution given the asymmetric growth of bubbles and spikes. The detailed parameters are shown in Tables \ref{tbl:2dSimpara} and \ref{tbl:3dSimpara} in \ref{app:simpara}. In the simulations, $Re_p$ is varied by adjusting $\mu$ (and $\kappa$ to keep $Pr=1$) and fixing other parameters.

The initial density follows an error function profile in the vertical direction \cite{wei2012late}:
\begin{gather}
\rho(x,y,z) = 0.5\{1+\text{erf}[Y_v\;z+\xi]\}(\rho_h-\rho_l)+\rho_l,
\end{gather}
where $Y_v=170$ is the slope coefficient, $\xi$ is the initial density perturbation in the form of:
\begin{gather}
\xi(x) =\mathcal{A}cos(\frac{2\pi x}{\lambda}) \;\text{in 2D, and}\\
\xi(x, y) =\mathcal{A}[cos(\frac{2\pi x}{\lambda})+cos(\frac{2\pi y}{\lambda})] \;\text{in 3D}. 
\end{gather}
Here, $\mathcal{A}=0.5$ is the amplitude of $\xi$. For single-mode RTI, the perturbation wavelength is $\lambda=L_x$. The actual amplitude of the density profile is $\mathcal{A}/Y_v$ \cite{wei2012late}. Figure  \ref{fig:IC} illustrates the initial condition along the center line ($x=L_x/2$ and $y=L_y/2$) from a simulation with $A=0.25$. 

Periodic boundary conditions are used in the horizontal directions. In the vertical direction, we have no-slip rigid walls for the velocity, a zero heat flux boundary condition for temperature, and a hydrostatic condition $dP/dz|_{z=0, L_z}=-\rho g$ for pressure.

\section{RESULTS}
\label{sec:result}
\begin{figure}
\caption*{\raggedright\hspace{0.3cm} \hspace{0cm}$Re_p=100$\hspace{0.8cm}$Re_p=1000$\hspace{0.6cm}$Re_p=20000$ }
\vspace{-0.2cm}
\centering
\begin{subfigure}{0.04\textwidth}
\includegraphics[width=\textwidth]{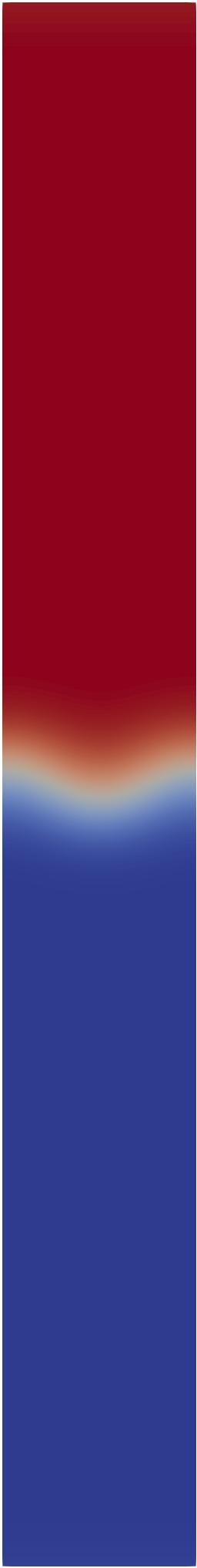}
\caption{$\tau{=}2$}
\end{subfigure}
\begin{subfigure}{0.04\textwidth}
\includegraphics[width=\textwidth]{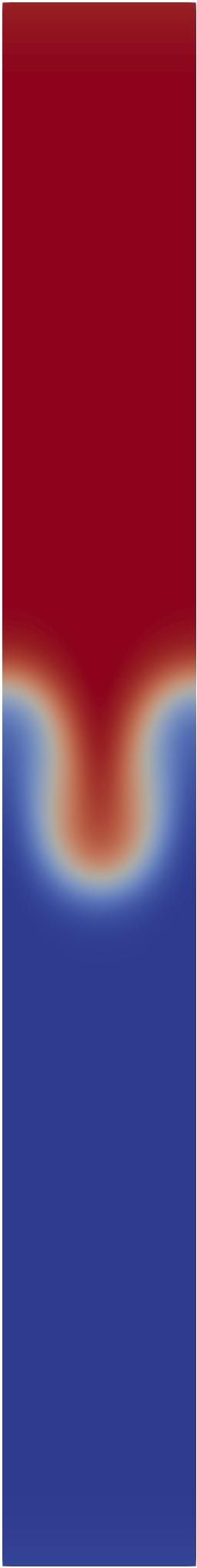}
\caption{$\tau{=}4$}
\end{subfigure}
\begin{subfigure}{0.04\textwidth}
\includegraphics[width=\textwidth]{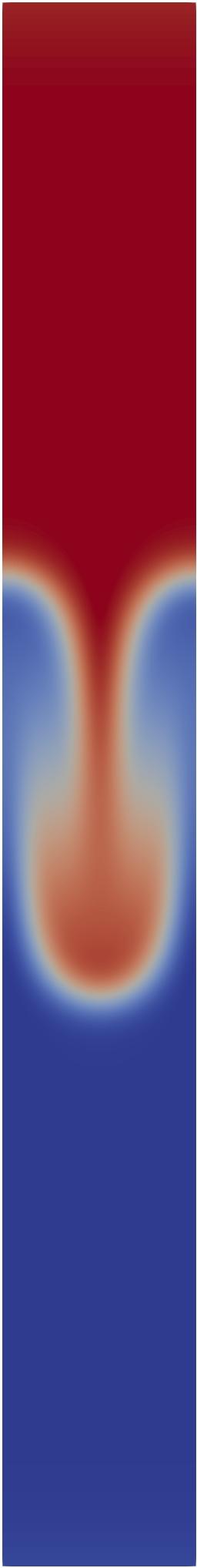}
\caption{$\tau{=}6$}
\label{}
\end{subfigure}
\begin{subfigure}{0.04\textwidth}
\includegraphics[width=\textwidth]{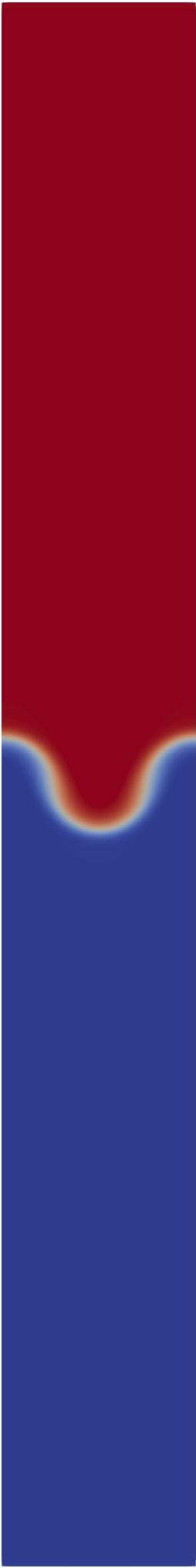}
\caption{$\tau{=}2$}
\end{subfigure}
\begin{subfigure}{0.04\textwidth}
\includegraphics[width=\textwidth]{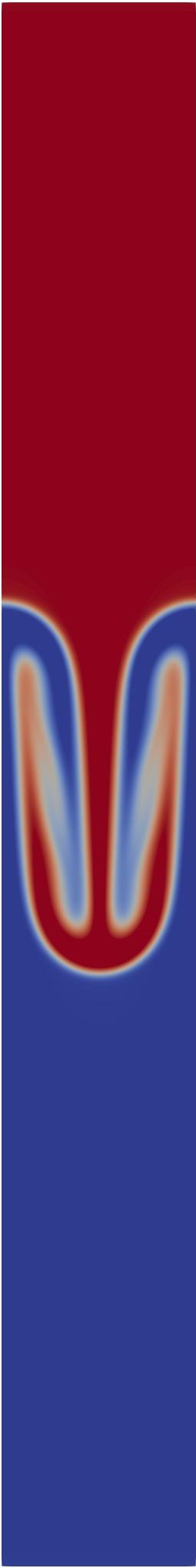}
\caption{$\tau{=}4$}
\end{subfigure}
\begin{subfigure}{0.04\textwidth}
\includegraphics[width=\textwidth]{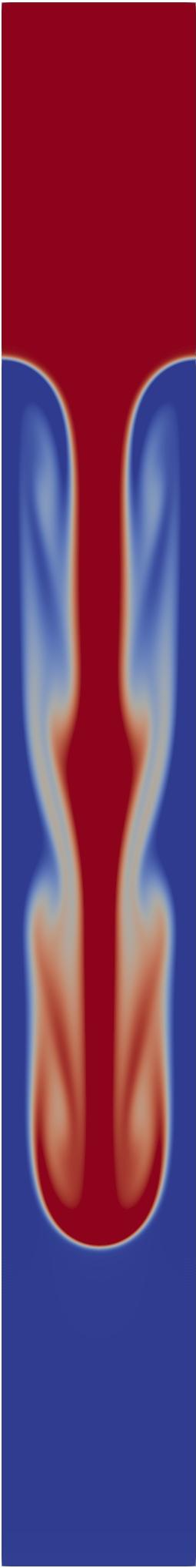}
\caption{$\tau{=}6$}
\label{}
\end{subfigure}
\begin{subfigure}{0.04\textwidth}
\includegraphics[width=\textwidth]{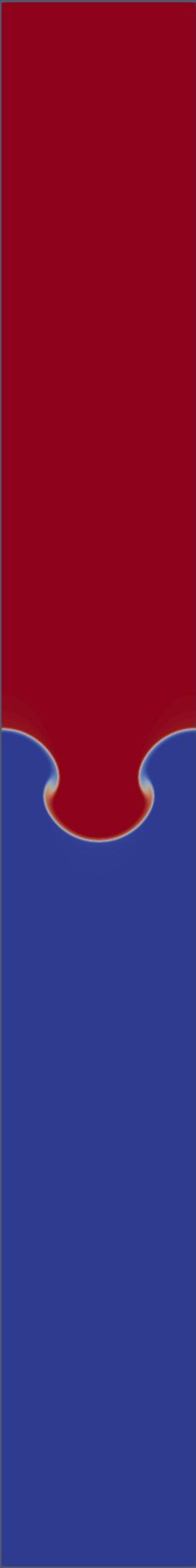}
\caption{$\tau{=}2$}
\label{}
\end{subfigure}
\begin{subfigure}{0.04\textwidth}
\includegraphics[width=\textwidth]{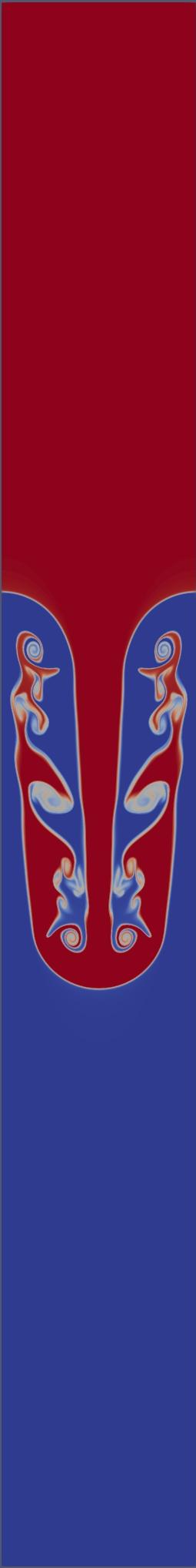}
\caption{$\tau{=}4$}
\end{subfigure}
\begin{subfigure}{0.04\textwidth}
\includegraphics[width=\textwidth]{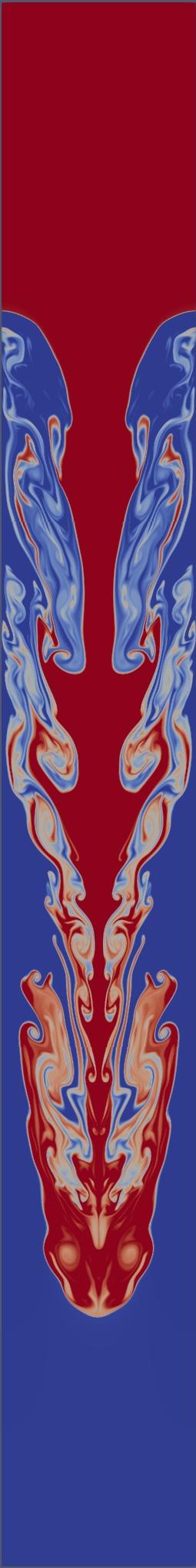}
\caption{$\tau{=}6$}
\label{}
\end{subfigure}
\begin{subfigure}{0.04\textwidth}
\includegraphics[width=\textwidth]{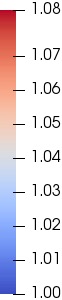}
\vspace{-1cm}
\caption*{}
\label{}
\end{subfigure}
\renewcommand{\figurename}{FIG.}
\caption{2D density $\rho$ visualizations at $A=0.04$ for three typical $Re_p=$ 100, 1000, and 20000. (a-c) show results of $Re_p$ = 100 at $\tau=2$, 4, and 6. (d-f) are results of $Re_p$ = 1000 at $\tau=2$, 4, and 6. (g-i) show results of $Re_p$ = 20000 at $\tau=2$, 4, and 6. The plots show more vortical structures are generated at higher $Re_p$}
\label{fig:2dRhoVisDiffRe}
\end{figure}

\begin{figure*}
\centering
\begin{subfigure}{0.45\textwidth}
\includegraphics[width=2.2 in]{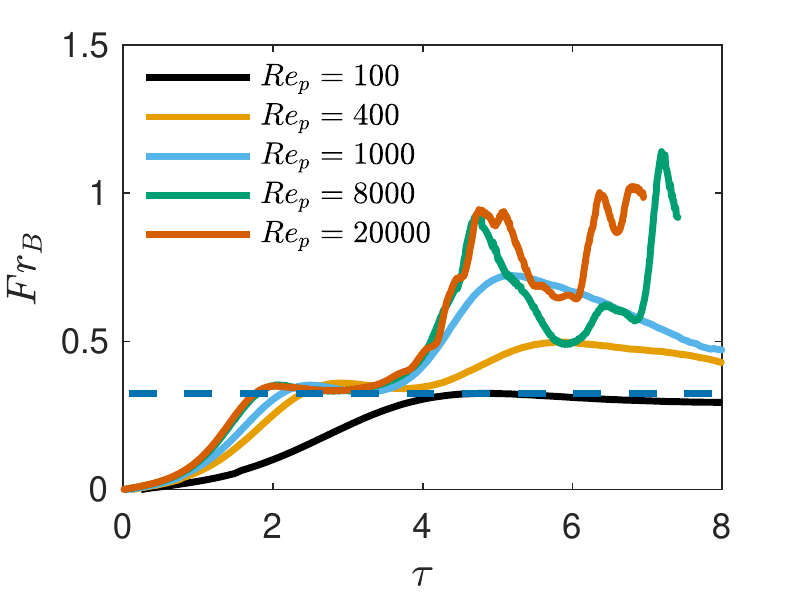}
\end{subfigure}
\begin{subfigure}{0.45\textwidth}
\includegraphics[width=2.2 in]{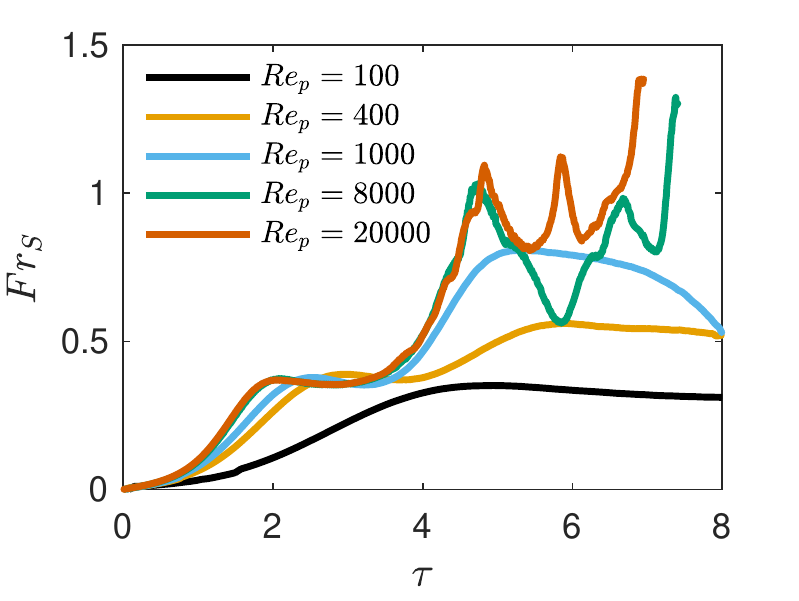}
\end{subfigure}
\\
\vspace{-0.15cm}
\begin{subfigure}{0.45\textwidth}
\includegraphics[width=2.2 in]{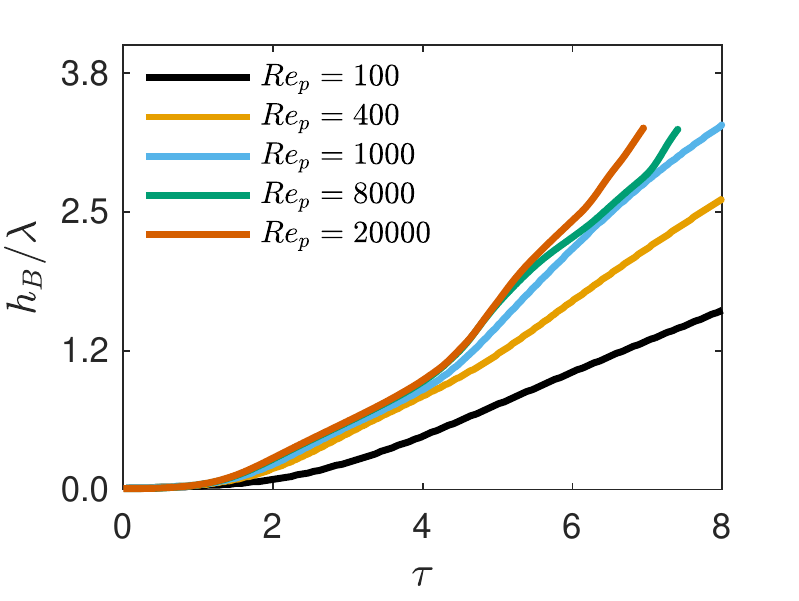}
\end{subfigure}
\begin{subfigure}{0.45\textwidth}
\includegraphics[width=2.2 in]{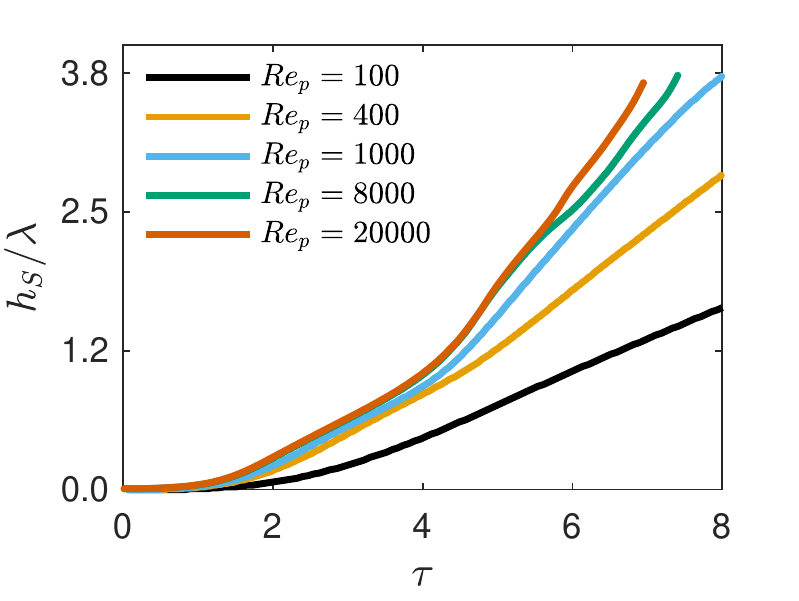}
\end{subfigure}
\renewcommand{\figurename}{FIG.}
\caption{Effects of $Re_p$ on the bubble velocity $Fr_B$ (top left), the spike velocity $Fr_B$ (top right), the bubble front location $h_B$ (bottom left), and the spike front location $h_S$ (bottom right) in 2D RTI at $A=0.04$. The dashed line shows the potential model prediction of $Fr_B = \sqrt {1/(3\pi)}$. $h_{B/S}$ is measured relative to the initial interface position $z_0=0.5L_z$. \\ The plots show that the bubble and spike growth (both velocities and heights) are similar at low $A$ and $Re_p$, however they experience asymmetric growth at late times at the highest $Re_p$, even at low $A=0.04$ cases shown here. A sustainable re-acceleration stage is observed at high $Re_p$. The results are similar to the incompressible RTI results in Ref. \cite{wei2012late}. Note that, $\tau$ here is smaller than the definition used in Ramaprabhu et al. \cite{ramaprabhu2006limits,ramaprabhu2012late} by a factor $\sqrt{2\pi}$ ($\tau=6$ here corresponds to time $\approx15$ in \cite{ramaprabhu2006limits,ramaprabhu2012late}).}
\label{fig:bubSpeedDiffRe}
\end{figure*}

We will now show a systematic analysis of the influence of $Re_p$ and $A$ in both 2D and 3D compressible RTI. At low $A=0.04$, we observe bubble re-acceleration that is temporally persistent when $Re_p$ exceeds a threshold value ($Re_p=6000$ in 2D  and $Re_p=400$ in 3D).
We also observe the emergence of asymmetry in the bubble and spike development (height and velocity) at late times, even at the lowest $A=0.04$ when $Re_p$ is high. At moderate $Re_p$ below the threshold, re-acceleration is only transient and the bubble eventually decelerates as in \cite{ramaprabhu2012late}. However, the deceleration does not seem to stop at a constant bubble velocity. Even at $Re_p=100$, when the re-acceleration does not occur, the constant velocity growth is not maintained for long times as the bubble eventually exhibits slight deceleration. At a fixed $Re_p$, increasing $A$ above a threshold suppresses the bubble re-acceleration. However, the $A$ threshold value increases for higher $Re_p$, suggesting that re-acceleration can be maintained as $A\to 1$ if the $Re_p\to\infty$ limit is taken first. We find that the effect of $Re_p$ and $A$ in 3D is qualitatively similar to 2D except that re-acceleration is easier to attain in 3D, requiring lower $Re_p$ threshold values. An analysis of the vorticity dynamics shows that bubble re-acceleration is indeed due to the vortices propagating into the bubble tip \cite{betti2006bubble,ramaprabhu2012late}.  

\subsection{Effect of Perturbation Reynolds Number}
\label{sec:reynolds}

Figure \ref{fig:2dRhoVisDiffRe} presents the density visualizations for $A=0.04$ at different $Re_p$. The corresponding bubble and spike development (velocities and heights) is plotted in Fig. \ref{fig:bubSpeedDiffRe}. The bubble (spike) height, $h_{B/S}$, is defined by the location (relative to the initial interface) of the maximum vertical density gradient, $\partial_z\rho|_{x=0}$ ($\partial_z\rho|_{x=L_x/2}$), along the line $x=0$ ($x=L_x/2$). The bubble/spike velocity, $Fr_{B/S}$, is calculated from the vertical velocity at that location. 

Figure \ref{fig:bubSpeedDiffRe} shows three main trends for $Fr_B$ development, with the plots for $Re_p=100$, 1000, and 20000 being representative of these trends. During the linear stage at early times, $Fr_B$ increases exponentially for all cases. When the bubble height amplitude exceeds the nonlinear criterion ($h_B\approx$ 0.1 $\lambda$), the bubble velocity becomes saturated. The saturation value agrees with Goncharov's ``terminal velocity'' from potential theory \cite{goncharov2002analytical}.  However, in none of the simulations, this constant velocity growth is fully maintained to the end of the simulation. At $Re_p=100$, after reaching the saturation value predicted by Goncharov slightly after $\tau=4$, $Fr_B$ starts to slowly decay. When $Re_p$ exceeds 100, the bubble accelerates beyond this saturation value at later times in the deep nonlinear phase ($h_B\gtrsim \lambda$). The bubble velocity at $Re_p=1000$ reaches more than twice the ``terminal velocity'' before eventually decaying. At yet higher $Re_p$ ($Re_p=8000$ and 20000), the bubble undergoes further re-acceleration ($\tau \approx$ 6) in the deep nonlinear phase. This stage was the onset of the so-called ``chaotic development'' stage in \cite{wei2012late}, where the bubble rises with a mean constant acceleration \cite{wei2012late}. However, our fully compressible simulations are more expensive than in \cite{wei2012late} and we were not able to advance to sufficiently late times to observe a clear chaotic development stage. 

While a higher $Re_p$ also leads to a larger growth rate during the early linear stage, $\gamma = \sqrt{Akg - 4 \nu k^2 \gamma}$  (see Fig. \ref{fig:bubSpeedDiffRe}), such a correlation does not need to imply a causal relation between the viscous effects during the linear stage and their effects on the bubble speed in the deep nonlinear stage. Indeed, from Fig. \ref{fig:bubSpeedDiffRe}, we observe that despite the different linear growth rates for different $\nu$, all cases plateau to the same value at the end of the linear stage. As shown in Figs. \ref{fig:2dRhoVisDiffRe}(c, f, i), the density visualization at $\tau=6$ suggests more vortices are generated by KHI at higher $Re_p$. This suggests that the mechanism by which viscosity affects the growth during the deep nonlinear stage is different in its nature and is primarily involved in the damping of the generation and advection of vorticity into the bubble tip. Section \ref{sec:small-scale-diss} will show that the re-acceleration in the deep nonlinear phase is indeed due to the vortices propagating into the bubble tip.

The time evolution of the spike velocity and height (right panels in Fig. \ref{fig:bubSpeedDiffRe}) is very similar to that of the bubble, especially at early times, $\tau < 4$. At later times, the spike velocity, especially at high $Re_p>1000$, is larger than the corresponding bubble velocity, while the results of low $Re_p=100$ are almost identical. This highlights the asymmetry in the development between bubbles and spikes even at very low $A=0.04$.

The main conclusions from analyzing $Re_p$ dependence are that a) at sufficiently large $Re_p$, the enhancement in bubble velocity beyond the ``terminal'' value is sustained and does not decrease at later times as had been previously observed in lower-resolution simulations \cite{ramaprabhu2012late} and b) even at lower $Re_p$, when the re-acceleration seems to fail or is not achieved over the duration of our simulations, the bubble velocity does not maintain a constant value but decays instead at late times. Note that we do not observe re-acceleration at lower $Re_p$, even when simulated over times longer than those reported here (using longer domains). We attribute this to a lack of vortices in the bubble tip, which will be shown to be the reason for bubble re-acceleration in Section \ref{sec:small-scale-diss}.

\subsection{The Effects of Atwood Number}
\label{sec:atwood}
\begin{figure*}
\caption*{\raggedright\hspace{1.6cm}$A=0.04$\hspace{2.4cm}$A=0.25$\hspace{2.5cm}$A=0.6$\hspace{2.7cm}$A=0.8$ }
\vspace{-0.4cm}
\centering
\begin{subfigure}{0.04\textwidth}
\includegraphics[width=\textwidth]{densityVis/re20000t=2.jpg}
\caption{$\tau{=}2$}
\end{subfigure}
\begin{subfigure}{0.04\textwidth}
\includegraphics[width=\textwidth]{densityVis/re20000t=4.jpg}
\caption{$\tau{=}4$}
\end{subfigure}
\begin{subfigure}{0.04\textwidth}
\includegraphics[width=\textwidth]{densityVis/re20000t=6.jpg}
\caption{$\tau{=}6$}
\end{subfigure}
\begin{subfigure}{0.04\textwidth}
\includegraphics[width=\textwidth]{densityVis/1_083_colorbar.jpg}
\vspace{-1cm}
\caption*{}
\label{}
\end{subfigure}
\hspace{0.5cm}
\begin{subfigure}{0.04\textwidth}
\includegraphics[width=\textwidth]{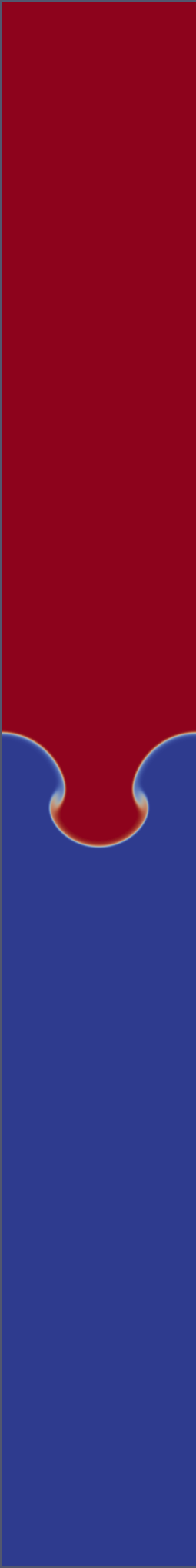}
\caption{$\tau{=}2$}
\label{}
\end{subfigure}
\begin{subfigure}{0.04\textwidth}
\includegraphics[width=\textwidth]{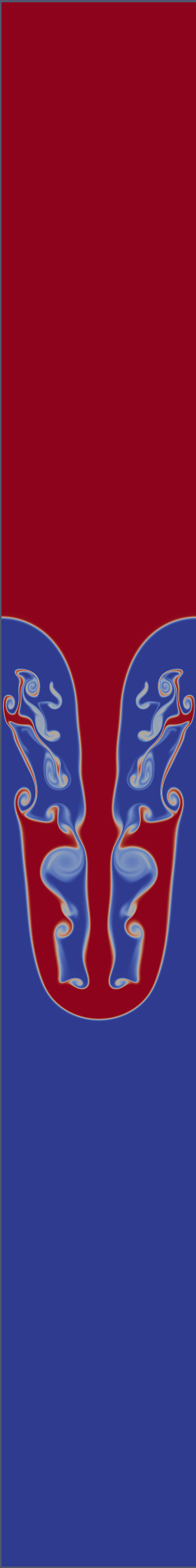}
\caption{$\tau{=}4$}
\label{}
\end{subfigure}
\begin{subfigure}{0.04\textwidth}
\includegraphics[width=\textwidth]{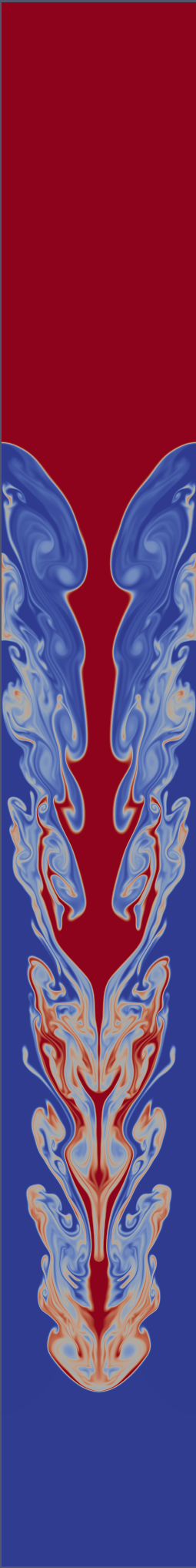}
\caption{$\tau{=}6$}
\label{}
\end{subfigure}
\begin{subfigure}{0.04\textwidth}
\includegraphics[width=\textwidth]{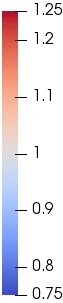}
\vspace{-1cm}
\caption*{}
\label{}
\end{subfigure}
\hspace{0.5cm}
\begin{subfigure}{0.04\textwidth}
\includegraphics[width=\textwidth]{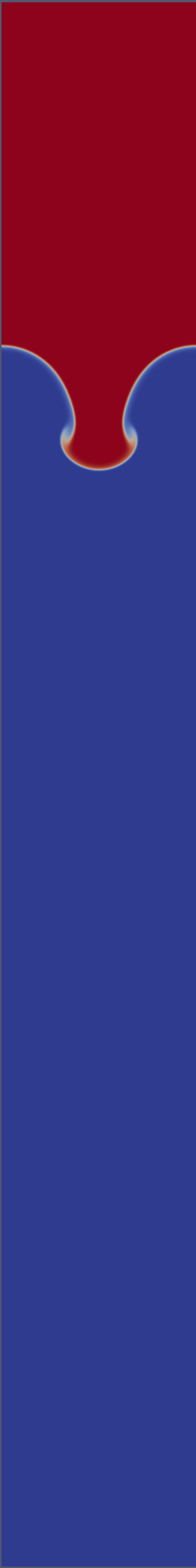}
\caption{$\tau{=}2$}
\label{}
\end{subfigure}
\begin{subfigure}{0.04\textwidth}
\includegraphics[width=\textwidth]{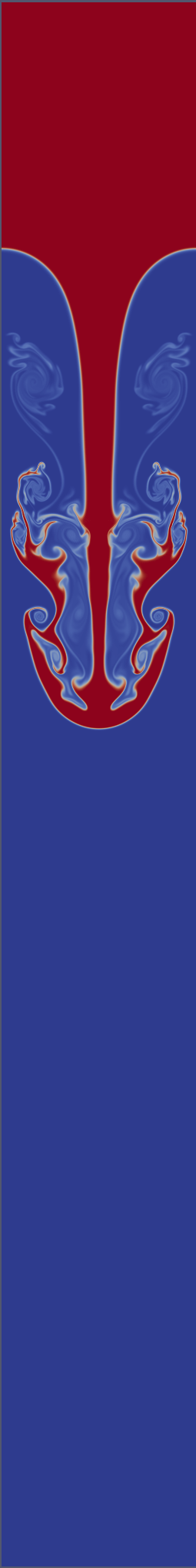}
\caption{$\tau{=}4$}
\label{}
\end{subfigure}
\begin{subfigure}{0.04\textwidth}
\includegraphics[width=\textwidth]{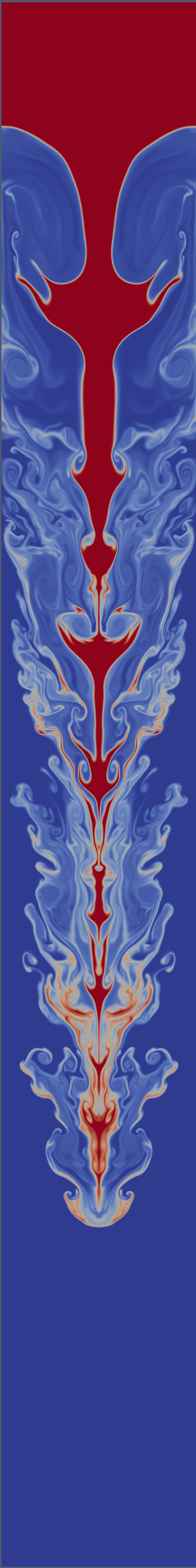}
\caption{$\tau{=}6$}
\label{}
\end{subfigure}
\begin{subfigure}{0.04\textwidth}
\includegraphics[width=\textwidth]{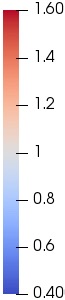}
\vspace{-1cm}
\caption*{}
\label{}
\end{subfigure}
\hspace{0.5cm}
\begin{subfigure}{0.04\textwidth}
\includegraphics[width=\textwidth]{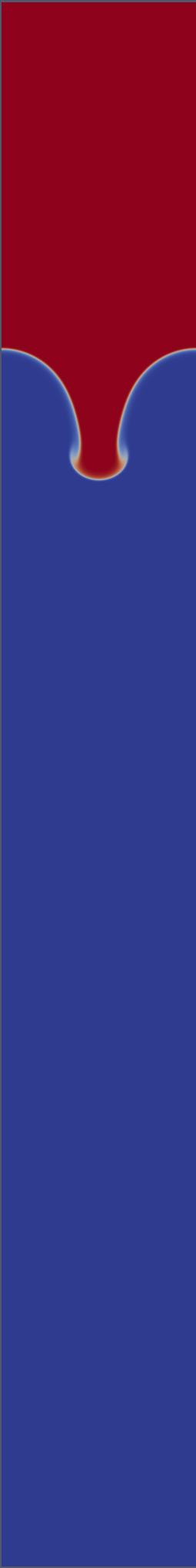}
\caption{$\tau{=}2$}
\label{}
\end{subfigure}
\begin{subfigure}{0.04\textwidth}
\includegraphics[width=\textwidth]{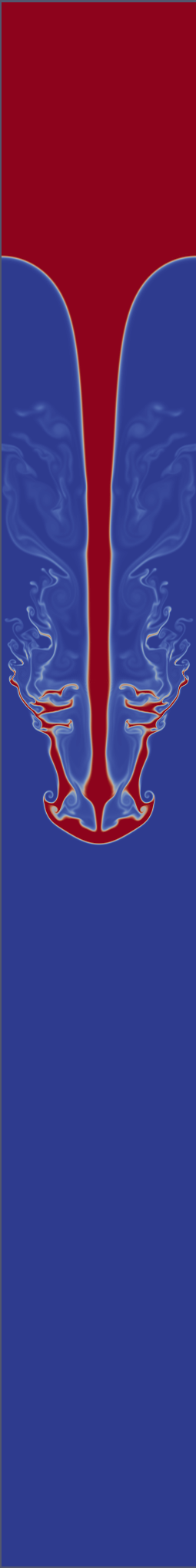}
\caption{$\tau{=}4$}
\label{}
\end{subfigure}
\begin{subfigure}{0.04\textwidth}
\includegraphics[width=\textwidth]{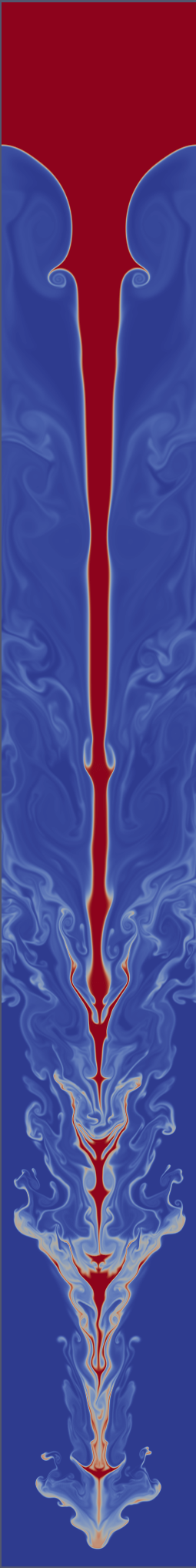}
\caption{$\tau{=}6$}
\label{}
\end{subfigure}
\begin{subfigure}{0.04\textwidth}
\includegraphics[width=\textwidth]{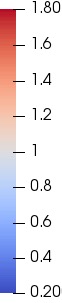}
\vspace{-1cm}
\caption*{}
\label{}
\end{subfigure}
\renewcommand{\figurename}{FIG.}
\caption{2D density $\rho$ visualizations at $Re_p=20000$ for $A=$ 0.04, 0.25, 0.6, and 0.8.  (a-c) show the results of $A$ = 0.04 at $\tau=2$, 4, and 6. (d-f) are the results of $A$ = 0.25 at $\tau=2$, 4, and 6. (g-i) show the results of $A$ = 0.6 at $\tau=2$, 4, and 6. Note the $A=0.04$ and 0.25 simulations have an initial interface position at $z_0=0.5L_z$, while the $A=0.6$ and 0.8 simulations' initial interface position is at $z_0=0.75L_z$. Also note that in the $A=0.8$ simulation, the spike reaches the bottom soon after $\tau=6$.}
\label{fig:2dRhoVisDiffA}
\end{figure*}

\begin{figure*}
\centering
\begin{subfigure}{0.45\textwidth}
\includegraphics[width=2.2 in]{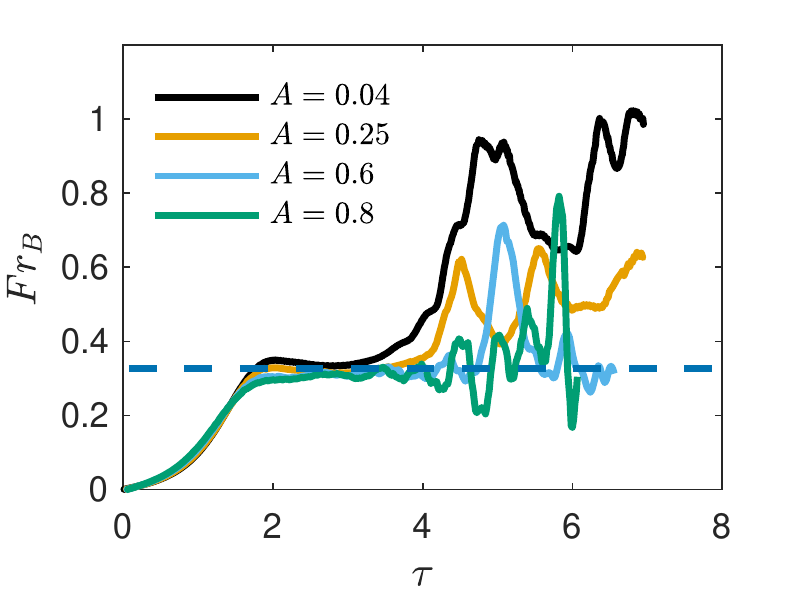}
\end{subfigure}
\begin{subfigure}{0.45\textwidth}
\includegraphics[width=2.2 in]{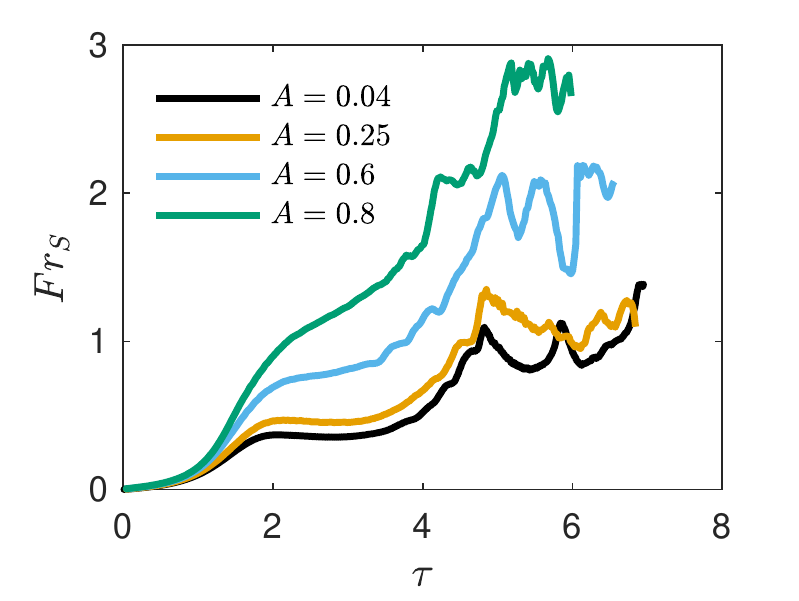}
\end{subfigure}
\\
\vspace{-0.15cm}
\begin{subfigure}{0.45\textwidth}
\includegraphics[width=2.2 in]{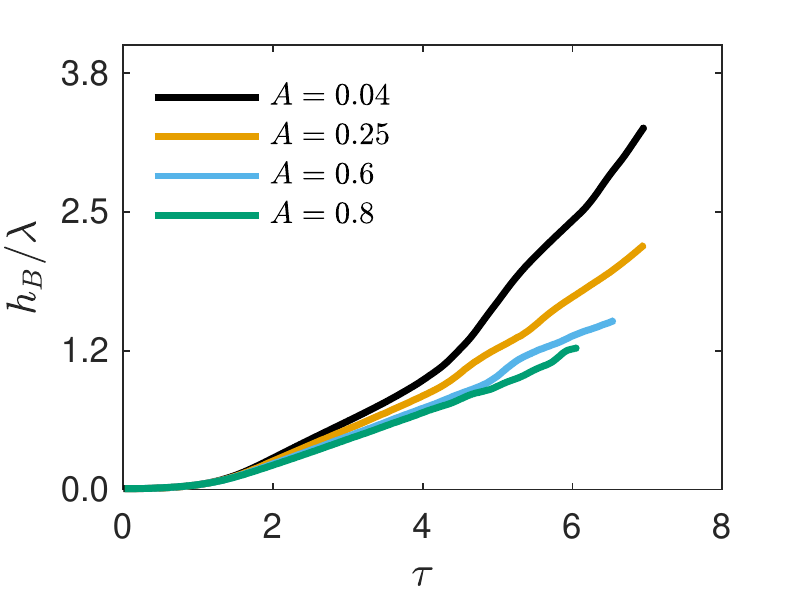}
\end{subfigure}
\begin{subfigure}{0.45\textwidth}
\includegraphics[width=2.2 in]{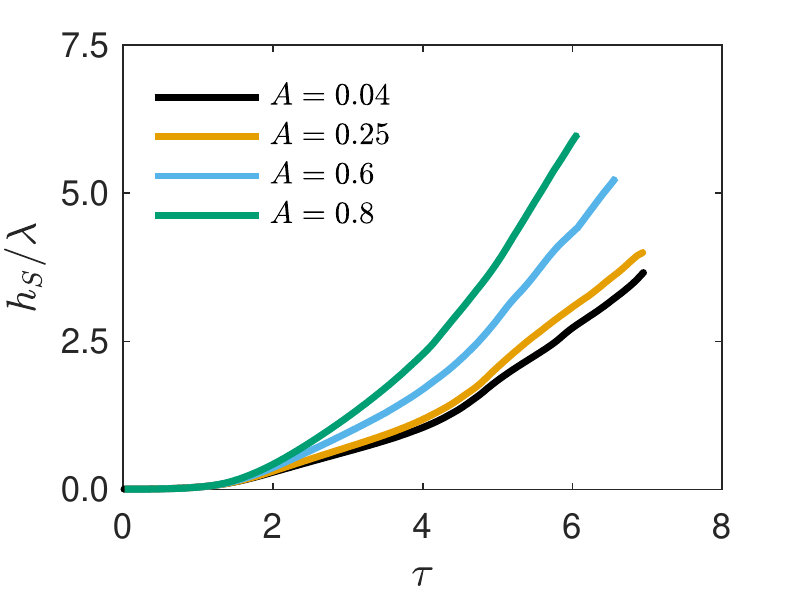}
\end{subfigure}
\renewcommand{\figurename}{FIG.}
\caption{Effects of $A$ on the bubble velocity $Fr_B$ (top left), the spike velocity $Fr_B$ (top right), the bubble front location $h_B$ (bottom left), and the spike front location $h_S$ (bottom right) in 2D RTI at $Re_p$ = 20000. The dashed line shows the potential model prediction of $Fr_B = \sqrt {1/(3\pi)}$. $h_{B/S}$ is measured relative to the initial interface position $z_0=0.5L_z$ at $A=0.04$ and $0.25$ and $z_0=0.75L_z$ at $A=0.6$ and $0.8$. The plots show the absence of the bubble re-acceleration and the asymmetric development of bubbles and spikes at high $A$. Note that, $\tau$ here is smaller than the definition used in Ramaprabhu et al. \cite{ramaprabhu2006limits,ramaprabhu2012late} by a factor $\sqrt{2\pi}$ ($\tau=6$ here corresponds to time $\approx15$ in \cite{ramaprabhu2006limits,ramaprabhu2012late}).}
\label{fig:bubSpeedDiffA}
\end{figure*}

Figure \ref{fig:2dRhoVisDiffA} presents density visualizations for $Re_p=20000$ at different $A$. The corresponding bubble and spike development (velocities and heights) are shown in Fig. \ref{fig:bubSpeedDiffA}. The simulations with higher $A$ ($A=0.6$ and 0.8) end sooner due to the spike approaching the bottom wall in less time. We also note that the spike for the $A=0.8$, $Re_p=20000$ case exhibits very slight asymmetry at the latest times, close to the wall. In the four cases, the time evolution of the bubble velocity at different $A$ is similar during the linear and early nonlinear stages ($h_B\lesssim \lambda$). At later times, varying $A$ leads to differing development trends. For the $A=0.04$ case, the bubble front undergoes re-acceleration and exhibits the putative onset of chaotic development as discussed in the previous subsection. For the $A=0.25$ case, the bubble front still undergoes re-acceleration but with a smaller magnitude. In the simulations with $A=$ 0.6 and 0.8, the bubble front experiences repeated unsustainable re-accelerations and its speed decays to temporarily low values at later times. A clear and sustainable re-acceleration at high $A$ is not observed over the time and $Re_p$ in these simulations. Due to the vortices transported towards the bubble tip, we anticipate that bubble velocity keeps fluctuating at later time instead of saturating at a constant velocity. In addition, the morphology of the layer becomes different than when the quasi-constant bubble velocity is first observed. Therefore, it is doubtful that the original ``terminal velocity'' has any relevance for the long time behavior. However, a more definitive statement on the bubble behavior at even later times requires taller domains which is beyond what we were able to perform in this study.

\begin{figure}
\centering
\includegraphics[width=2.2 in]{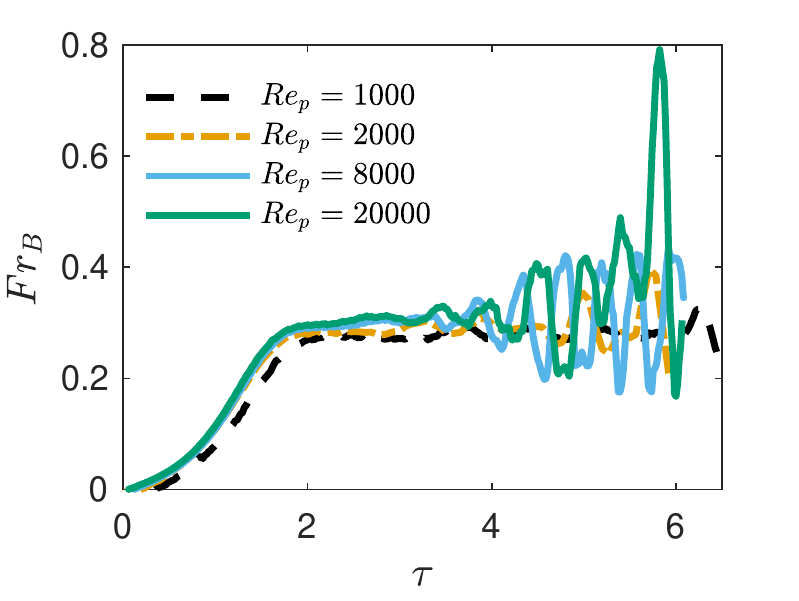}
\renewcommand{\figurename}{FIG.}
\caption{Time evolutions of bubble velocities at $A=0.8$ for different $Re_p$. The plot indicates that the fluctuation amplitude becomes stronger as $Re_p$ increases and that bubble re-acceleration is possible at sufficiently high $Re_p$.}
\label{fig:compareDiffReforA08}
\end{figure}

To better understand the bubble front behavior at high $A$, Figure  \ref{fig:compareDiffReforA08} compares the evolution of bubble velocities for $A=0.8$ at different $Re_p$. The plots show a clear trend for more intense fluctuations in the bubble velocity as $Re_p$ increases. This suggests that a sustainable re-acceleration regime can appear at arbitrarily high $A$ if $Re_p$ is large enough, although determining this with definitive certainty requires even higher resolution simulations beyond what we were able to perform in this study. 

Figure \ref{fig:bubSpeedDiffA} allows for comparing the bubble and spike velocities at one of the highest $Re_p=20000$ value considered in this study. The spike approaches free-fall behavior and experiences weaker resistive drag as $A$ increases from 0.04 to 0.8. For $A=0.04$ and 0.25, the spike reaches a quasi-constant velocity growth before re-accelerating at $\tau \approx 3.5$. After that, there are subsequent acceleration and deceleration phases around $Fr_S=1$, which might be indicative of the emergence of the ``chaotic growth'' regime, with mean quadratic growth. However, the simulations are not long enough to clearly identify this regime. Larger $A$ leads to larger deviations from the ``terminal velocity'' $\sqrt{\frac{2Ag}{(1 - A)3k}}$ obtained from potential flow theory \cite{goncharov2002analytical}, which has been well-known to be due to vortices generated near the spike tip. For the highest $A=0.8$, the spike speed never passes through a constant velocity stage.

The evolutions of bubble and spike heights are also plotted in Fig. \ref{fig:bubSpeedDiffA}. The bubble height development shows that $h_B$ increases faster as $A$ decreases. In contrast, the spike height development shows that $h_S$ increases faster as $A$ increases. The plots suggest that the asymmetric development between bubbles and spikes becomes more significant at larger $A$, as is well-known. For example, $h_S$ is about 4.7 times larger than $h_B$ for $A=0.8$ at $\tau\approx6$, while $h_S\approx1.12 h_B$ for $A=0.04$ at the same time.

The main conclusion from analyzing $A$ dependence is that increasing $A$ makes it more difficult for bubble speed to increase and persist above the ``terminal velocity'' value of potential flow theory. This is consistent with the findings of Ramaprabhu et al. \cite{ramaprabhu2012late}. However, the results in \cite{ramaprabhu2012late} showed an eventual deceleration back to the ``terminal velocity'' after a transient re-acceleration stage for all Atwood numbers, including low $A=0.005$ (see Figs. 7 (a) and (c) in Ref. \cite{ramaprabhu2012late}). In contrast, our results indicate that the bubble speed enhancement above the ``terminal'' value can be sustained regardless of $A$ if the $Re_p$ is sufficiently large. The differing results are most probably due to the difference in resolution and momentum conservation. The results reported here maintain symmetry and are at a significantly higher resolution than what was possible several years ago when \cite{ramaprabhu2012late} was conducted. Compared to the simulations in \cite{ramaprabhu2012late} our simulations show a clear and sustained bubble speed enhancement at $A=0.04$ and $0.25$. At higher $A>0.25$, the bubble velocity exhibits intermittent oscillations above the ``terminal'' value with an intensity that increases with increasing $Re_p$, suggesting that a clear sustained bubble speed enhancement is possible if $Re_p$ is sufficiently large. Future studies using higher resolution simulations with larger $Re_p$ are encouraged to verify the bubble's late-time behavior at high $A$, which is of particular relevance in ICF.

\subsection{3D Effects}
\label{sec:3deffects}

\begin{figure}
\centering
\caption*{\raggedright\hspace{0.6cm} $A=0.04$\hspace{3.0cm}$A=0.8$}
\vspace{-0.3cm}
\begin{subfigure}{0.08\textwidth}
\caption* {\footnotesize $Re_p{=}1000$}
\includegraphics[width=\textwidth]{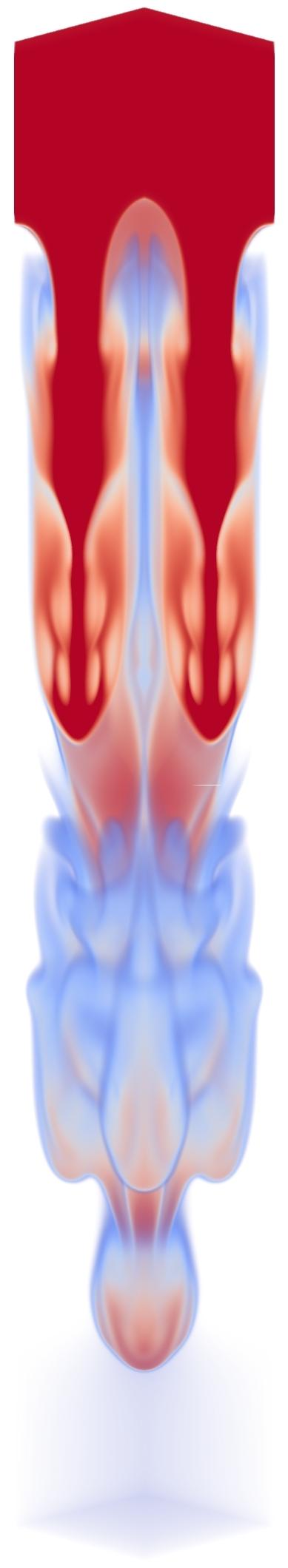}
\vspace{-1.4cm}
\caption {}
\end{subfigure}
\begin{subfigure}{0.08\textwidth}
\caption* {\footnotesize $Re_p{=}8000$}
\includegraphics[width=\textwidth]{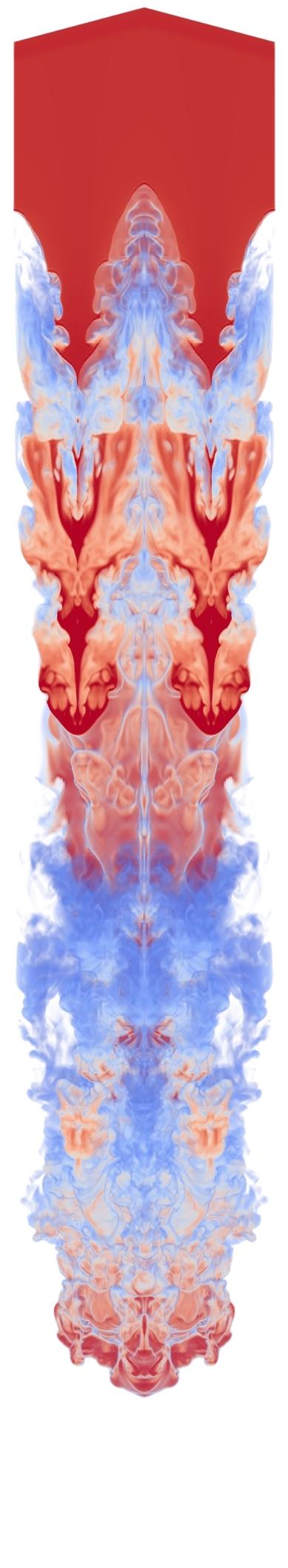}
\vspace{-1.4cm}
\caption {}
\end{subfigure}
\begin{subfigure}{0.04\textwidth}
\includegraphics[width=\textwidth]{densityVis/1_083_colorbar.jpg}
\vspace{-1cm}
\caption*{}
\end{subfigure}
\hspace{0.5 cm}
\begin{subfigure}{0.08\textwidth}
\caption* {\footnotesize $Re_p{=}1000$}
\includegraphics[width=\textwidth]{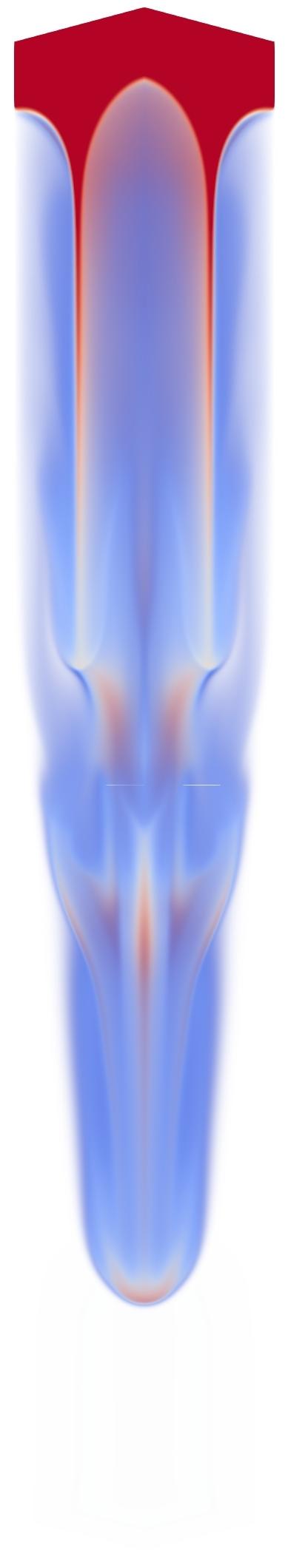}
\vspace{-1.4cm}
\caption {}
\end{subfigure}
\begin{subfigure}{0.08\textwidth}
\caption* {\footnotesize $Re_p{=}8000$}
\includegraphics[width=\textwidth]{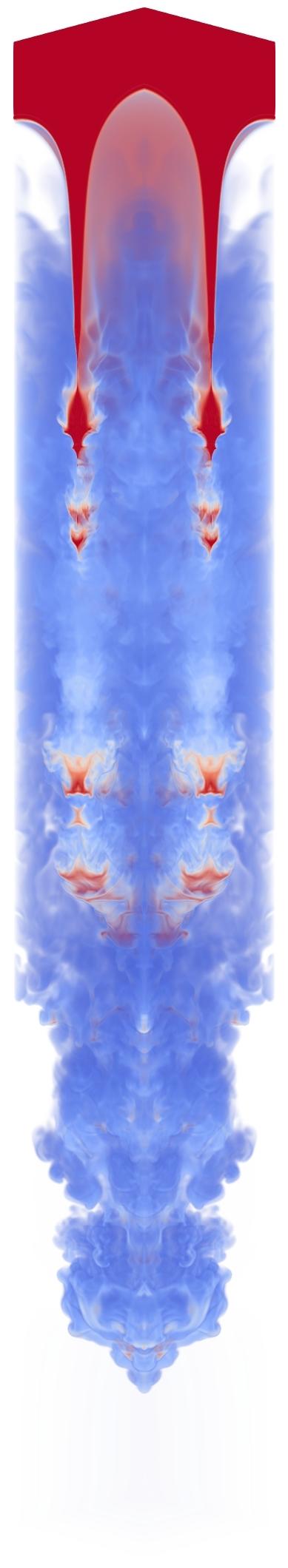}
\vspace{-1.4cm}
\caption {}
\end{subfigure}
\begin{subfigure}{0.04\textwidth}
\includegraphics[width=\textwidth]{densityVis/1_8_colorbar.jpg}
\vspace{-1cm}
\caption*{}
\end{subfigure}
\renewcommand{\figurename}{FIG.}
\caption{3D density $\rho$ visualizations at $\tau=5$. (a), (b) are results at $A=0.04$ for $Re_p=1000$ and 8000, respectively. (c), (d) are results at $A=0.8$ for $Re_p=1000$ and 8000, respectively. Note the initial interface position is $z_0=0.5L_z$ for $A=0.04$ and $z_0=0.75L_z$ for $A=0.8$. The plots show that 3D RTI generates more vortical structures than 2D RTI.}
\label{fig:3dRhoVis}
\end{figure}

\begin{figure*}
\centering
\begin{subfigure}{0.45\textwidth}
\includegraphics[width=2.2 in]{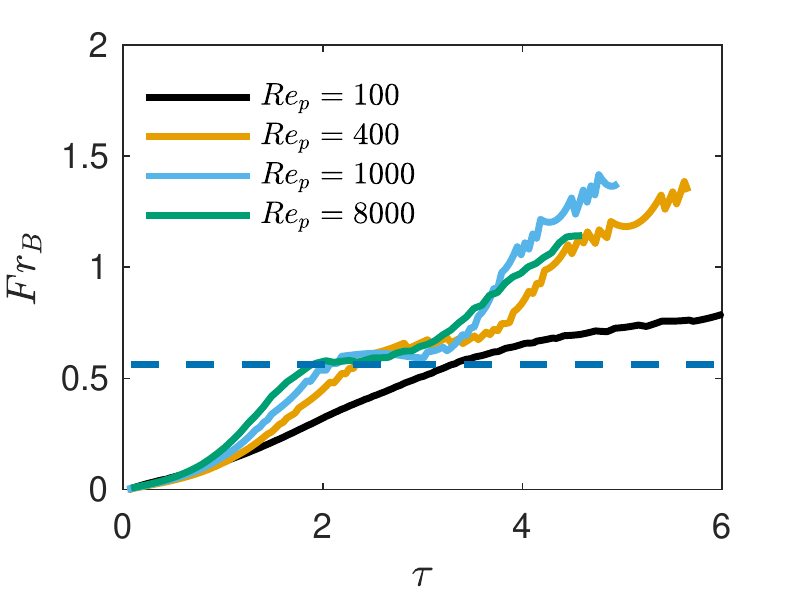}
\end{subfigure}
\begin{subfigure}{0.45\textwidth}
\includegraphics[width=2.2 in]{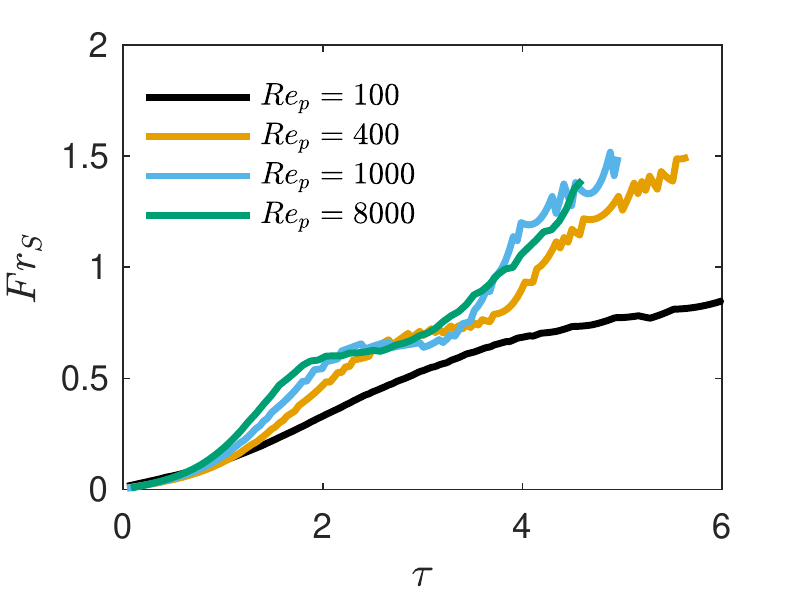}
\end{subfigure}
\renewcommand{\figurename}{FIG.}
\caption{Time evolutions of bubble (left panel) and spike (right panel) velocities at $A=0.04$ and different $Re_p$ in 3D RTI. The dashed line shows the potential model of $Fr_B = \sqrt {1/\pi}$. The plots show 3D RTI develop faster and is easier to re-accelerate than its 2D counterpart. Note that, $\tau$ here is smaller than the definition used in Ramaprabhu et al. \cite{ramaprabhu2006limits,ramaprabhu2012late} by a factor $\sqrt{2\pi}$ ($\tau=6$ here corresponds to time $\approx15$ in \cite{ramaprabhu2006limits,ramaprabhu2012late}).}
\label{fig:bubSpeedAt0043D}
\end{figure*}
  
\begin{figure*}
\centering
\begin{subfigure}{0.45\textwidth}
\includegraphics[width=2.2 in]{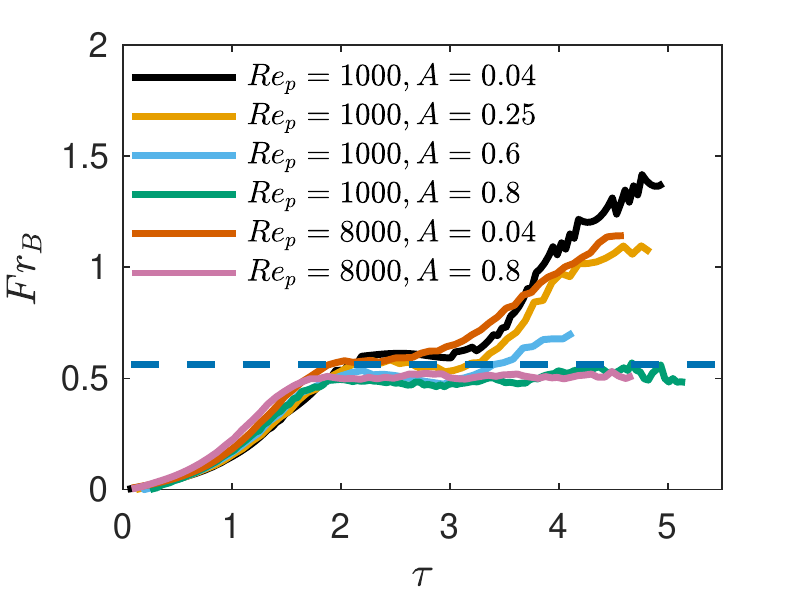}
\end{subfigure}
\begin{subfigure}{0.45\textwidth}
\includegraphics[width=2.2 in]{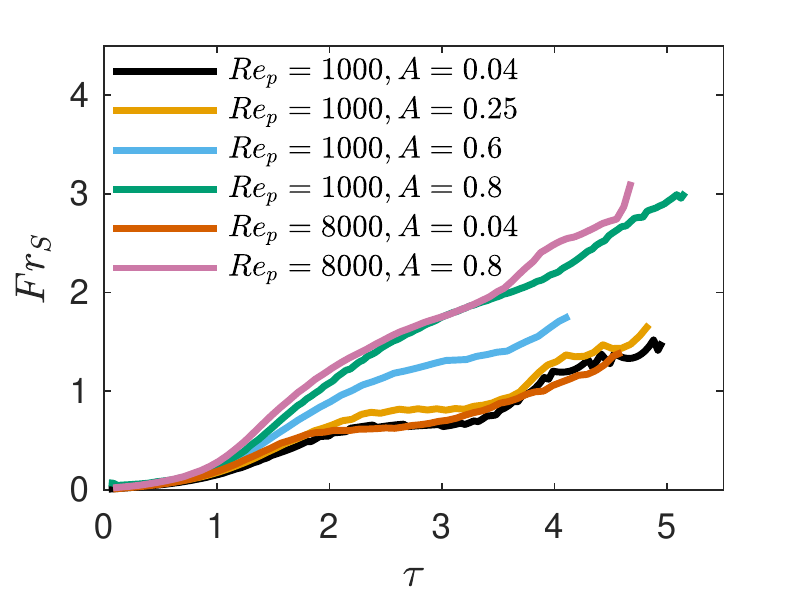}
\end{subfigure}
\renewcommand{\figurename}{FIG.}
\caption{Time evolutions of the bubble (left panel) and spike (right panel) velocities at different $A$ for two $Re_p$ in 3D RTI. The dashed line shows the potential model of $Fr_B = \sqrt {1/\pi}$. Note that, $\tau$ here is smaller than the definition used in Ramaprabhu et al. \cite{ramaprabhu2006limits,ramaprabhu2012late} by a factor $\sqrt{2\pi}$ ($\tau=6$ here corresponds to time $\approx15$ in \cite{ramaprabhu2006limits,ramaprabhu2012late}).}
\label{fig:bubSpeedDiffA3D}
\end{figure*}

We now report on 3D effects at different $Re_p$ and $A$. The main conclusion is that compared to 2D results, 3D bubbles develop faster and require smaller $Re_p$ threshold values (e.g. $Re_p=400$ at $A=0.04$) for re-acceleration.

Figure  \ref{fig:3dRhoVis} presents density visualizations at $\tau=5$. The bubble velocity for $A=0.04$ at different $Re_p$ is plotted in Fig. \ref{fig:bubSpeedAt0043D}. It is shown that the bubble velocity in all cases increases persistently without returning to the ``terminal velocity''  and does not exhibit the temporal fluctuations observed in 2D. Even at the lowest $Re_p=100$ and 400, the bubble never seems to pass through a constant velocity phase after the exponential growth of the linear stage. At $A=0.04$, the plots also show that the evolutions of the bubble and spike velocities are very similar, although they become asymmetric at high $Re_p$ at late times.

The absence of a constant velocity stage and larger growth rates in 3D are not surprising. Faster (properly normalized) growth rates in 3D than in 2D were already shown by previous studies in both single-mode and multi-mode RTI \cite{tryggvason1990computations,youngs1991three}. In 3D RTI, the assumption in potential theory is easier to violate than that in 2D due to  vortex stretching which significantly amplifies vorticity. As a result, the density contour in Fig. \ref{fig:3dRhoVis} indicates much more intense vorticity generation in 3D than in 2D, resulting in faster bubbles and spikes. In addition, vortex rings self-propagate faster than vortex pairs.

Overall, our results indicate weaker requirements for re-acceleration in 3D than in 2D. For example, at $Re_p$ = 1000, the bubble front can re-accelerate at $A\leq0.25$ (see Fig. \ref{fig:bubSpeedDiffA3D}), while, the bubble front in 2D exhibits only a weak re-acceleration followed by a deceleration at later times, even for $A$ = 0.04 at this $Re_p$ value (see Fig. \ref{fig:bubSpeedDiffRe}).  

\subsection{Re-acceleration Phase Diagram}
\label{sec:ra-excitation}
\begin{figure*}
\centering
\begin{subfigure}{0.45\textwidth}
\includegraphics[width=2.2 in]{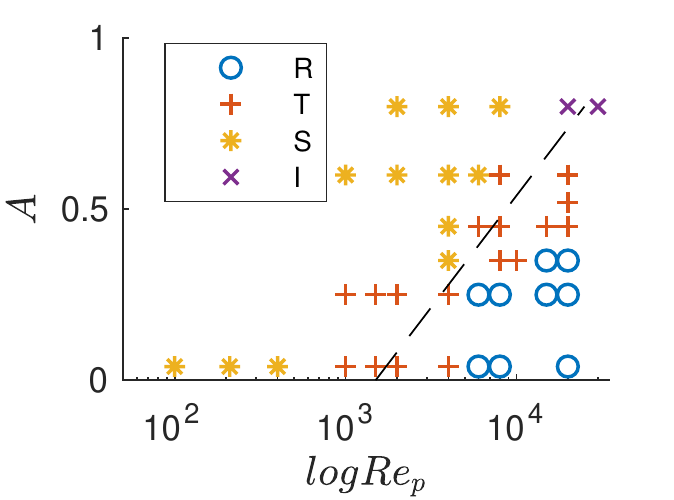}
\end{subfigure}
\begin{subfigure}{0.45\textwidth}
\includegraphics[width=2.2 in]{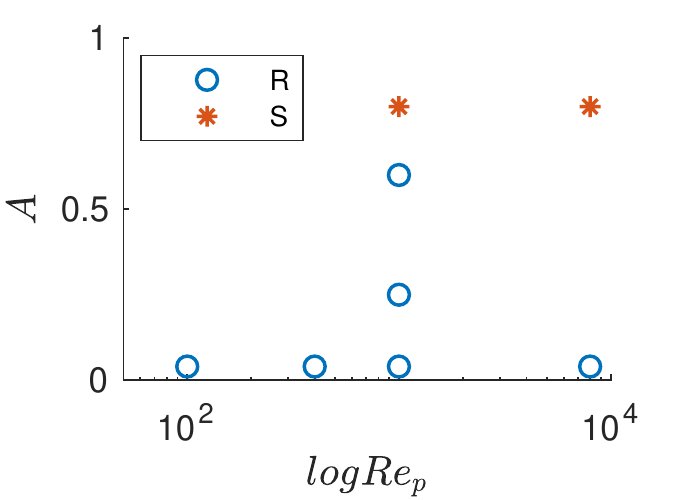}
\end{subfigure}
\renewcommand{\figurename}{FIG.}
\caption{Re-acceleration $A$ -- $Re_p$ phase diagram in 2D (left panel) and 3D (right panel) RTI. The late-time behavior of the bubble front is classified into 4 phases: 1) (R) denotes a robust re-acceleration exists at late times; 2)  (T)ransient means the bubble front re-acceleration is temporary and eventually decelerates at late times; 3) (S)aturation phase indicates that the bubble front saturates near the ``terminal velocity'' after the linear stage, with possible eventual decay at very long times; 4) (I)ntermittent phase means the bubble front velocity is characterized by intermittent large amplitude fluctuations without clear long time trend. 
}
\label{fig:exitCond}
\end{figure*}

The above results show the influence of $A$ and $Re_p$ on  re-acceleration. To further elucidate the threshold parameter values required for re-acceleration, the late-time behavior of the bubble front is classified into 4 phases in Fig. \ref{fig:exitCond}: robust re-acceleration (R), transient (T), saturation (S), and intermittent (I). A robust re-acceleration (R) phase means the bubble re-accelerates and eventually displays an onset of the chaotic development stage (e.g. $Re_p=20000$, $A=0.04$ in 2D); in the Transient (T) phase, a bubble re-accelerates but then decelerates  (e.g. $Re_p=1000$, $A=0.04$ and $Re_p=20000$, $A=0.6$ in 2D); in the  saturation (S) phase, re-acceleration is absent after the linear growth stage and the bubble front  saturates for some time near the theoretical value from potential theory or fluctuates slightly around it, with likely eventual decay at very late times  (e.g. $Re_p$=100, $A=0.04$ and $Re_p=8000$, $A=0.8$ in 2D). $Re_p=20000$ and 30000 at $A=0.8$ in 2D are considered in the intermittent phase (I) due to the large intermittent oscillations of $Fr_B$ above the ``terminal velocity'' value, with no clear late time trend. 

Figure  \ref{fig:exitCond} summarizes these findings in a phase diagram in $A$ and $Re_p$ space. The figure makes clear how the threshold $Re_p$ value for a re-acceleration increases with increasing $A$. In 2D RTI, 1) for $Re_p\le 400$, late-time RTI is in the (S)aturated phase. 2) For $1000\le Re_p \le 6000$,  late-time RTI is in the (T)ransient phase at small $A \le 0.25$. Increasing $A$ to larger values, re-acceleration is completely suppressed and late-time RTI is in the (S)aturated phase. 3) When $6000 \le Re_p \le 8000$, late-time RTI is in the (R)obust Re-acceleration phase when $A$ is small enough. Increasing $A$ at a constant $Re_p$, late-time RTI changes phase from (R) to (T) to (S). 4) For $Re_p\ge 20000$, the speed is highly intermittent and it is not clear whether or not the bubble will eventually re-accelerate at $A=0.8$.

Due to the limited computing resources, the dependence of re-acceleration on $A$ and $Re_p$ in 3D is less thoroughly investigated here. Compared to 2D RTI, a robust re-acceleration appears at smaller $Re_p$ and larger $A$, such as $Re_p$ = 1000, $A=0.25$ (note that the case of $Re_p$ =1000, $A=0.6$ is considered as (R)obust Re-acceleration phase here). The (T)ransient and (I)ntermittent phases from 2D RTI are not observed in 3D simulations.

\subsection{Discussion on vortical structures}
\label{sec:small-scale-diss}

Following the work of Betti and Sanz \cite{betti2006bubble} and Ramaprabhu et al. \cite{ramaprabhu2006limits,ramaprabhu2012late},
we also investigated the role of vorticity in driving re-acceleration at different $Re_p$ and $A$. The results indicate a strong correlation between vorticity and re-acceleration, and show that discrete vortices are mainly generated away from the initial centerline, at the interface between lighter fluid and spike. These vortices then propagate towards the bubble tip, resulting in the bubble re-acceleration, consistent with results from ablative RTI \cite{betti2006bubble,yan2016three,Zhangetal18pre}. 

\begin{figure*}
\caption*{\raggedright\hspace{2.8cm} $Re_p=100$\hspace{2.5cm} $Re_p=1000$\hspace{2.5cm} $Re_p=20000$}
\vspace{-0.3cm}
\centering
\begin{subfigure}{0.04\textwidth}
\includegraphics[scale=0.8,trim={1cm 1.3cm 0.8cm 1.4cm},clip]{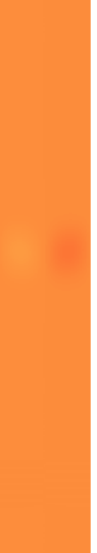}
\caption{$\tau{=}2$}
\end{subfigure}
\begin{subfigure}{0.04\textwidth}
\includegraphics[scale=0.8,trim={1cm 1.3cm 0.82cm 1.4cm},clip]{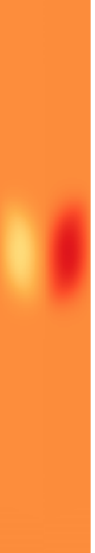}
\caption{$\tau{=}4$}
\end{subfigure}
\begin{subfigure}{0.04\textwidth}
\includegraphics[scale=0.8,trim={1cm 1.3cm 0.82cm 1.4cm},clip]{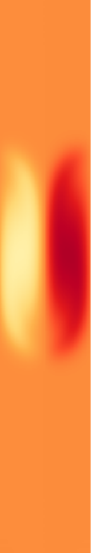}
\caption{$\tau{=}6$}
\label{}
\end{subfigure}
\begin{subfigure}{0.04\textwidth}
\includegraphics[scale=0.8,trim={2cm 1.3cm 0cm 1cm},clip]{vorticalPlots/re100_vorx_t6.pdf}
\caption*{}
\end{subfigure}
\hspace{1cm}
\begin{subfigure}{0.04\textwidth}
\includegraphics[scale=0.8,trim={1cm 1.3cm 0.95cm 1.4cm},clip]{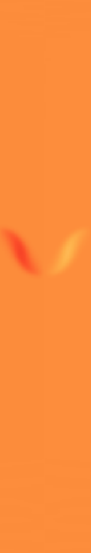}
\caption{$\tau{=}2$}
\end{subfigure}
\begin{subfigure}{0.04\textwidth}
\includegraphics[scale=0.8,trim={1cm 1.3cm 0.95cm 1.4cm},clip]{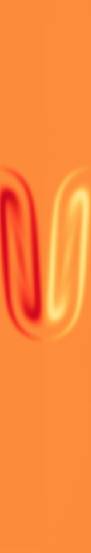}
\caption{$\tau{=}4$}
\end{subfigure}
\begin{subfigure}{0.04\textwidth}
\includegraphics[scale=0.8,trim={1cm 1.3cm 0.95cm 1.4cm},clip]{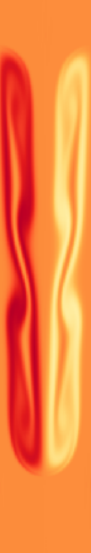}
\caption{$\tau{=}6$}
\label{}
\end{subfigure}
\begin{subfigure}{0.04\textwidth}
\includegraphics[scale=0.8,trim={2cm 1.3cm 0cm 1cm},clip]{vorticalPlots/re1000_vorx_t6.pdf}
\vspace{-0.5cm}
\caption*{}
\end{subfigure}
\hspace{1cm}
\begin{subfigure}{0.04\textwidth}
\includegraphics[scale=0.8,trim={1cm 1.3cm 0.95cm 1.4cm},clip]{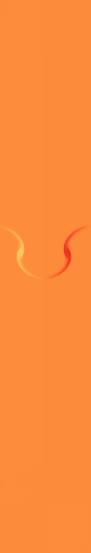}
\caption{$\tau{=}2$}
\label{}
\end{subfigure}
\begin{subfigure}{0.04\textwidth}
\includegraphics[scale=0.8,trim={1cm 1.3cm 0.95cm 1.4cm},clip]{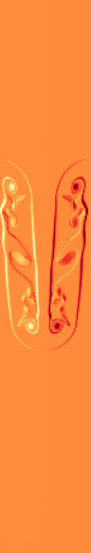}
\caption{$\tau{=}4$}
\end{subfigure}
\begin{subfigure}{0.04\textwidth}
\includegraphics[scale=0.8,trim={1cm 1.3cm 0.95cm 1.4cm},clip]{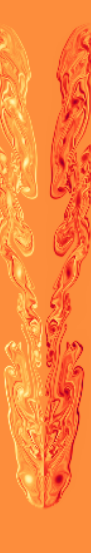}
\caption{$\tau{=}6$}
\label{}
\end{subfigure}
\begin{subfigure}{0.04\textwidth}
\includegraphics[scale=0.8,trim={2cm 1.3cm 0cm 1cm},clip]{vorticalPlots/re20000_vorx_t2.pdf}
\vspace{-0.5cm}
\caption*{}
\end{subfigure}
\renewcommand{\figurename}{FIG.}
\caption{Visualizations of 2D dimensionless vorticity $\omega^* = (\partial_z u_x - \partial_x u_z)/\sqrt{Ag/\lambda}$ at $A=0.04$ for different $Re_p$ = 100 (a-c), 1000 (d-f), and 20000 (g-i). The images are cropped vertically to save space. The plot shows larger amplitude of vorticity and stronger vortical motions at larger $Re_p$.}
\label{fig:2DvorxDiffRe}
\end{figure*}

Figure \ref{fig:2DvorxDiffRe} shows visualizations of dimensionless vorticity at $A=0.04$ for $Re_p$ = 100, 1000, and 20000 in 2D. We showed above how at $\tau=4$, the bubble in the $Re_p= 20000$ case starts to re-accelerate while that in the $Re_p=100$ case  does not exceed the ``terminal velocity'' from the potential flow theory (see Fig. \ref{fig:bubSpeedDiffRe}). Correspondingly, Figure  \ref{fig:2DvorxDiffRe}(h) shows complex vortical motions at $Re_p=20000$, while vortices in the $Re_p=100$ case are absent and vorticity intensity is $\approx 10\times$ lower. At $\tau=6$, Figure  \ref{fig:2DvorxDiffRe}(i) indicates abundant vortices inside the bubble tip at $Re_p=20000$. As we show below and discussed in \cite{betti2006bubble,ramaprabhu2006limits,ramaprabhu2012late,wei2012late}, these vortices are generated at the spike's interface as it penetrates into the lighter fluid. When the bubble symmetry (around the vertical axis) is maintained long enough and the viscosity is small enough so that the vortices do not dissipate, they then propagate into the bubble tip. The induced vortical velocity brings in purer (less mixed) lighter fluid from the interior, which increases the local Atwood number near the bubble tip, as well as exerts a direct centrifugal force, resulting in re-acceleration. The strength of the vortices inside the bubble tip fluctuates in time which, as we shall show below, correlates very well with the fluctuations in the bubble speed at late times. At intermediate $Re_p=1000$, Figure \ref{fig:2DvorxDiffRe}(f) shows that the vortices are weaker than that in $Re_p=20000$. Compared to the $Re_p=20000$ case, the $Re_p=1000$ flow shows less robust vorticity generation and a significant reduction in the strength and ubiquity of vortices which can propagate into the bubble to sustain a robust re-acceleration. The bubble velocity at $Re_p=1000$ eventually decays.
For $Re_p=100$, the bubble velocity does not reach values larger than the potential flow model because of the lack of vortices in the bubble tip, as shown in Fig. \ref{fig:2DvorxDiffRe}(c).

This phenomenology carries over to 3D single-mode RT where we have 3D vortex rings instead of 2D vortices generated and propagated toward the bubble tip (see Fig. \ref{fig:3dRhoVis}). These 3D vortex structures can get amplified due to stretching, thereby exerting an even stronger effect on bubble re-acceleration. Due to the axial symmetry, the induced advection due to vortex stretching is in the vertical direction. This has also been observed in 3D ablative RTI in \cite{yan2016three}. It is also consistent with our findings above where bubble re-acceleration is more pronounced in 3D than in 2D.

\begin{figure*}
\caption*{\raggedright\hspace{0.8cm} $A=0.04$\hspace{3.0cm} $A=0.25$\hspace{3.0cm} $A=0.6$\hspace{3.0cm} $A=0.8$}
\vspace{-0.3cm}
\centering
\begin{subfigure}{0.04\textwidth}
\includegraphics[scale=0.8,trim={1cm 0.5cm 0.95cm 0.6cm},clip]{vorticalPlots/re20000_vorx_t2.pdf}
\caption{$\tau{=}2$}
\label{}
\end{subfigure}
\begin{subfigure}{0.04\textwidth}
\includegraphics[scale=0.8,trim={1cm 0.5cm 0.95cm 0.6cm},clip]{vorticalPlots/re20000_vorx_t4.pdf}
\caption{$\tau{=}4$}
\end{subfigure}
\begin{subfigure}{0.04\textwidth}
\includegraphics[scale=0.8,trim={1cm 0.5cm 0.95cm 0.6cm},clip]{vorticalPlots/re20000_vorx_t6.pdf}
\caption{$\tau{=}6$}
\label{}
\end{subfigure}
\begin{subfigure}{0.04\textwidth}
\includegraphics[scale=0.8,trim={2cm 0.5cm 0cm 0.6cm},clip]{vorticalPlots/re20000_vorx_t2.pdf}
\caption*{}
\end{subfigure}
\hspace{1cm}
\begin{subfigure}{0.04\textwidth}
\includegraphics[scale=0.8,trim={1cm 0.5cm 0.95cm 0.6cm},clip]{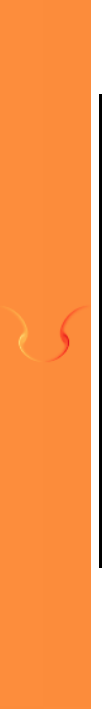}
\caption{$\tau{=}2$}
\label{}
\end{subfigure}
\begin{subfigure}{0.04\textwidth}
\includegraphics[scale=0.8,trim={1cm 0.5cm 0.95cm 0.6cm},clip]{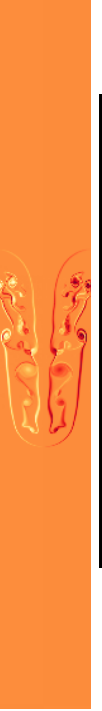}
\caption{$\tau{=}4$}
\end{subfigure}
\begin{subfigure}{0.04\textwidth}
\includegraphics[scale=0.8,trim={1cm 0.5cm 0.95cm 0.6cm},clip]{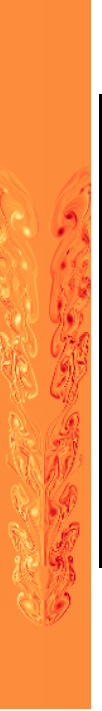}
\caption{$\tau{=}6$}
\label{}
\end{subfigure}
\begin{subfigure}{0.04\textwidth}
\includegraphics[scale=0.8,trim={2cm 0.5cm 0cm 0.6cm},clip]{vorticalPlots/a025_vorx_t6.pdf}
\caption*{}
\end{subfigure}
\hspace{1cm}
\begin{subfigure}{0.04\textwidth}
\includegraphics[scale=0.8,trim={1cm 0.5cm 0.95cm 0.6cm},clip]{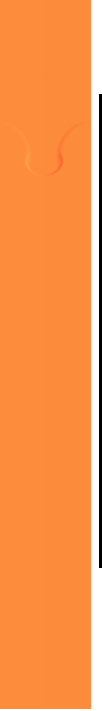}
\caption{$\tau{=}2$}
\label{}
\end{subfigure}
\begin{subfigure}{0.04\textwidth}
\includegraphics[scale=0.8,trim={1cm 0.5cm 0.95cm 0.6cm},clip]{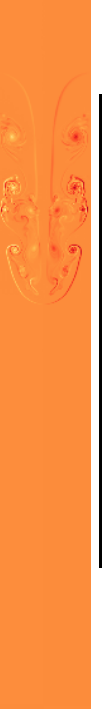}
\caption{$\tau{=}4$}
\end{subfigure}
\begin{subfigure}{0.04\textwidth}
\includegraphics[scale=0.8,trim={1cm 0.5cm 0.95cm 0.6cm},clip]{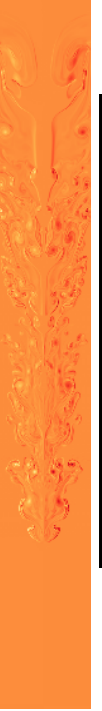}
\caption{$\tau{=}6$}
\label{}
\end{subfigure}
\begin{subfigure}{0.04\textwidth}
\includegraphics[scale=0.8,trim={2cm 0.5cm 0cm 0.6cm},clip]{vorticalPlots/a06_vorx_t6.pdf}
\vspace{-0.5cm}
\caption*{}
\end{subfigure}
\hspace{1cm}
\begin{subfigure}{0.04\textwidth}
\includegraphics[scale=0.8,trim={1cm 0.5cm 0.95cm 0.6cm},clip]{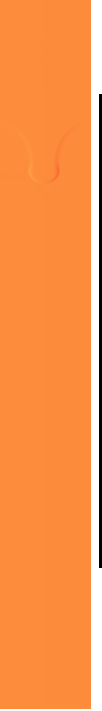}
\caption{$\tau{=}2$}
\label{}
\end{subfigure}
\begin{subfigure}{0.04\textwidth}
\includegraphics[scale=0.8,trim={1cm 0.5cm 0.95cm 0.6cm},clip]{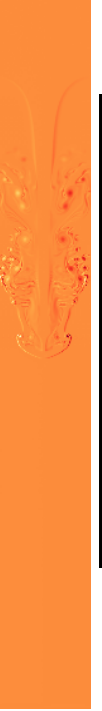}
\caption{$\tau{=}4$}
\end{subfigure}
\begin{subfigure}{0.04\textwidth}
\includegraphics[scale=0.8,trim={1cm 0.5cm 0.95cm 0.6cm},clip]{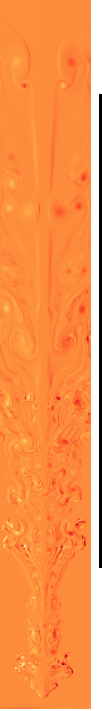}
\caption{$\tau{=}6$}
\label{}
\end{subfigure}
\begin{subfigure}{0.04\textwidth}
\includegraphics[scale=0.8,trim={2cm 0.5cm 0cm 0.cm},clip]{vorticalPlots/a08_vorx_t6.pdf}
\caption*{}
\end{subfigure}
\renewcommand{\figurename}{FIG.}
\caption{Visualizations of 2D dimensionless vorticity $\omega^* = (\partial_z u_x - \partial_x u_z)/\sqrt{Ag/\lambda}$ at $Re_p=20000$ for $A=0.04$ (a-c), 0.25 (d-f), 0.6 (g-i), 0.8 (j-l). Note the $A=0.04$ and 0.25 simulations have an initial interface position at $z=0.5L_z$, while the $A=0.6$ and 0.8 simulations' initial interface location is at $z=0.75L_z$. The images are cropped vertically to save space. Note the maximum amplitude of vorticity at higher $A$ is larger.}
\label{fig:2DvorxDiffA}
\end{figure*}

We now investigate why increasing $A$ tends to suppress re-acceleration. Figure  \ref{fig:2DvorxDiffA} presents the dimensionless vorticity visualizations for $Re_p=20000$ at different $A$. Remember that the spike velocity approaches free-fall behavior as $A\to1$, when the lighter fluid approaches vacuum. Since the spike growth increases relative to that of the bubble as $A$ increases (at any fixed $\tau$), the visualizations in Fig. \ref{fig:2DvorxDiffA} show that individual vortices have to travel longer distances before entering the bubble tip region. Thus, due to the increased spike velocity, the region with largest shear and, consequently, largest Kelvin-Helmholtz roll-up effect occurs closer to the spike tip, away from the initial (interface) centerline and bubble tip. This also correlates with the largest baroclinic vorticity production, similar to Ref. \cite{wieland2018}, where the stratification independent baroclinic torque was shown to dominate vorticity production. For the vortices generated below the initial centerline to reach the bubble tip region, two conditions must occur \cite{wei2012late}: a) the vortices need to move in the vertical direction faster than the bubble tip and b) they have to preserve their structure for sufficiently long times. 

The first condition can be satisfied if the symmetry around the (vertical) bubble axis, which is physically required in single-mode RTI due to momentum conservation, is maintained numerically. Maintaining the bubble symmetry  is important for re-acceleration since in that case vortices interact coherently (constructive interference) to induce a maximal vertical advective velocity to self-propagate. Otherwise, in the absence of symmetry, the resultant vertical advection from vortices is not as effective due to advection in the horizontal direction and possible destructive interference. Since the mean horizontal velocity is zero, a loss of symmetry would also imply interactions with other vortices, further hindering the vertical motions. Our simulations maintain this symmetry. 

The second condition requires that viscosity has to be small enough so that vortices do not dissipate or weaken enough to be displaced horizontally by other vortices, as they travel towards the bubble tip. Therefore, despite an increase in vorticity intensity, at fixed $Re_p$, increasing $A$ reduces the number of vortices entering the bubble tip, which prevents vorticity from aiding the bubble growth.

At $\tau=4$, the bubble starts to re-accelerate for $A\le0.25$ (see Fig. \ref{fig:bubSpeedDiffA}). In contrast, at $A\ge0.6$, the vortex rings need to travel a longer distance to affect the bubble dynamics, since the spike develops much faster at large Atwood numbers. For a fixed $Re_p$, the longer distance  traversed by vortices results in their dissipative attenuation before reaching the bubble tip. As a consequence, for $A\ge0.6$, Figures \ref{fig:2DvorxDiffA}(e, k) show that fewer vortices reach the bubble front.  At $\tau=6$, abundant vortices are present inside the bubble at $A\le0.25$ to sustain a robust re-acceleration, while the vortices at $A\ge0.6$ are less prevalent, which leads to the intermittent fluctuations in the bubble speed.

\begin{figure}
\centering
\begin{subfigure}{0.1\textwidth}
\includegraphics[width=\textwidth,trim={0.cm 5.3cm  .cm 5cm},clip]{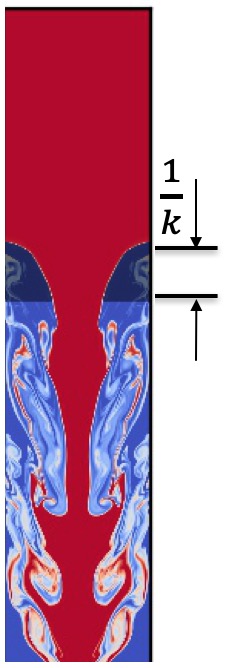}
\vspace{-0.5cm}
\caption {}
\end{subfigure}
\hspace{1cm}
\begin{subfigure}{0.1235\textwidth}
\includegraphics[width=\textwidth,trim={0.cm 5.3cm  .cm 5cm},clip]{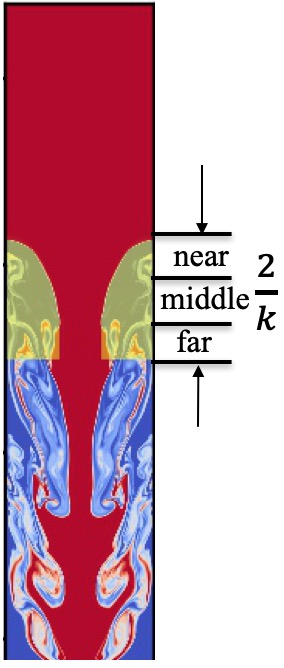}
\vspace{-0.5cm}
\caption {}
\end{subfigure}
\renewcommand{\figurename}{FIG.}
\caption{Plots illustrating the regions used in calculating vorticity $\omega_0 = \frac{\int_V\rho |\bm \omega|^2dV}{2\int_V\rho dV}$, where $V$ is the volume inside the bubble tip. $k=2\pi/l_x$ is the wavenumber. (a) The gray region is used in Fig. \ref{fig:omegaVelCorr}, which has a vertical distance of $1/k$.
(b) The yellow region is composed of three disjoint regions (near, middle and far regions). Each region has a vertical distance of $2/(3k)$. The three disjoint regions are used in Fig. \ref{fig:omega3Regions}.}
\label{fig:vor_explain}
\end{figure}

\begin{figure*}
\centering
\begin{subfigure}{0.45\textwidth}
\includegraphics[width=2.2 in]{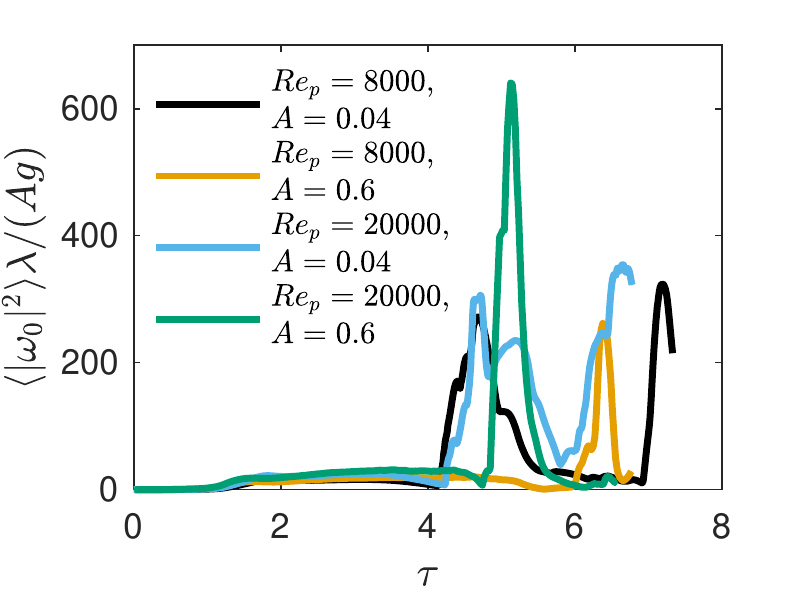}
\end{subfigure}
\begin{subfigure}{0.45\textwidth}
\includegraphics[width=2.2 in]{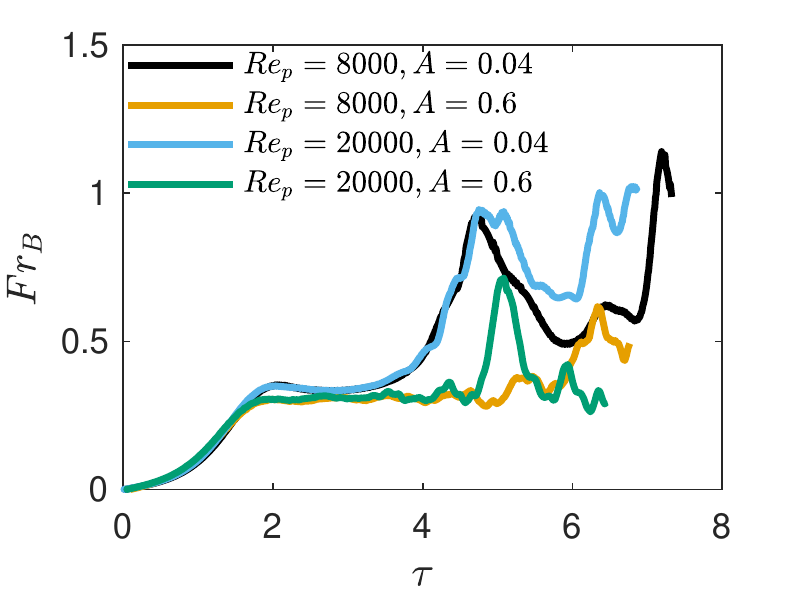}
\end{subfigure}
\renewcommand{\figurename}{FIG.}
\caption{Comparison between the spatially-averaged normalized vorticity $\langle|\omega_0|^2\rangle\lambda/(Ag)$ inside the bubble and the bubble velocity $Fr_B$ for 2D simulations. The plots show that there is a strong correlation between vorticity and bubble velocity in the nonlinear stage of the RTI growth.}
\label{fig:omegaVelCorr}
\end{figure*}

\begin{figure*}
\centering
\begin{subfigure}{0.45\textwidth}
\includegraphics[width=2.2 in]{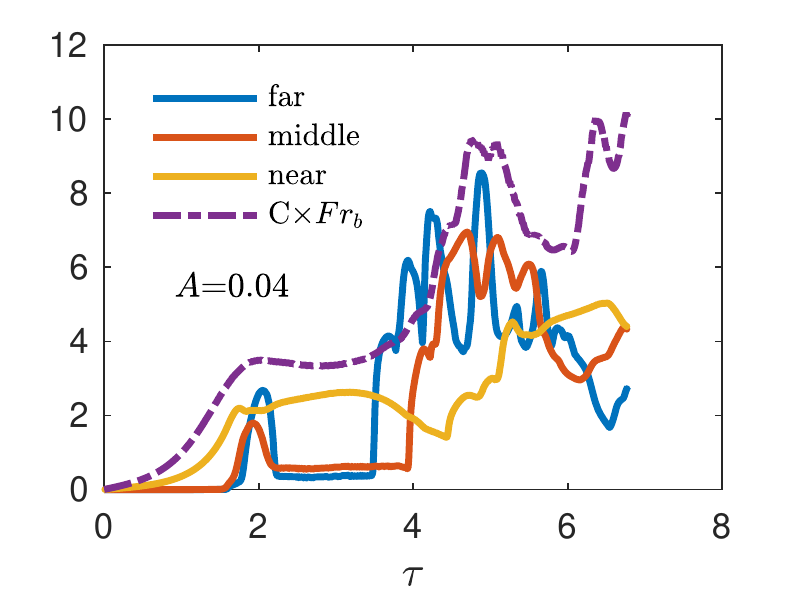}
\end{subfigure}
\begin{subfigure}{0.45\textwidth}
\includegraphics[width=2.2 in]{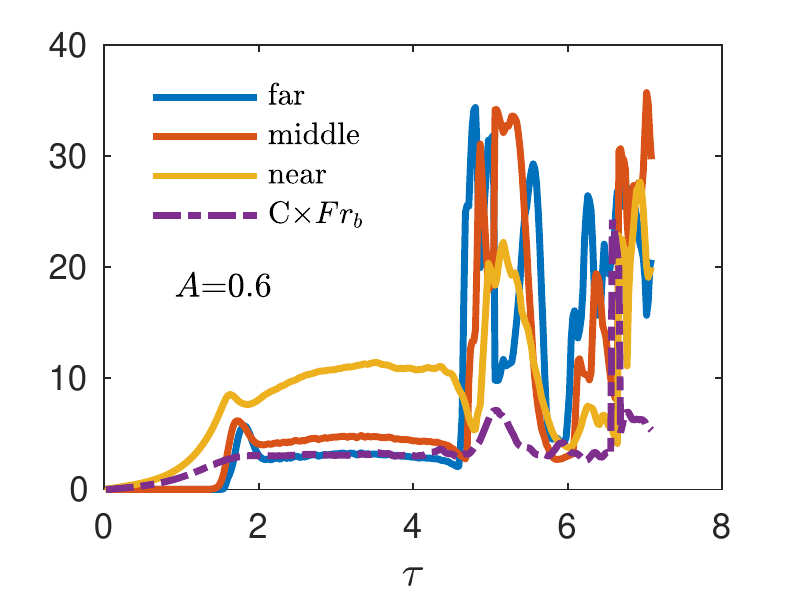}
\end{subfigure}
\renewcommand{\figurename}{FIG.}
\caption{Time evolutions of averaged vorticity $\omega_0$ within the three disjoint regions inside the bubble  from two 2D $Re_p=20000$ simulations. The regions are explained in Fig. \ref{fig:vor_explain}(b). The vertical distance for each region is $2/(3k)$. The dashed line shows $Fr_b$ multiplied by a constant C = 10 for better comparison. Left (right) panel shows $A=0.04$ ($A=0.6$) results. The results indicate vortices are advected from the bottom towards the bubble tip, and leads to  bubble re-acceleration at low $A=0.04$. At high $A=0.6$, the vortices are also advected from the bottom towards the bubble and leads to the fluctuations in speed around $\tau=6$.} 
\label{fig:omega3Regions}
\end{figure*}

To quantify the effect of vortices on bubble re-acceleration, the spatial average of vorticity behind a bubble tip $\omega_0$ (vertical length $1/k$, where $k$ is the wavenumber) is calculated. Here, $\omega_0 = \frac{\int_V\rho |\bm \omega|^2dV}{2\int_V\rho dV}$, ${\bm \omega} = {\bm \nabla\times u}$, and $V$ is the volume behind the bubble tip \cite{yan2016three} (the gray region in Fig. \ref{fig:vor_explain}(a)). The density weighted vorticity is used following \cite{betti2006bubble,yan2016three} since vortices in the heavier fluid have a larger momentum and exert a larger centrifugal force on the bubble (density weighting is more important at higher $A$). A comparison between the time evolution of $\langle|\omega_0|^2\rangle\lambda/(Ag)$ and $Fr_B$ is presented in Fig. \ref{fig:omegaVelCorr}. The plots indicate a strong correlation between vorticity and re-acceleration. The plateau in $\langle|\omega_0|^2\rangle\lambda/(Ag)$ from $\tau\approx1.5$ to 4 corresponds to the potential stage in the bubble velocity plots. For the $A$ = 0.04 simulations, $\langle|\omega_0|^2\rangle\lambda/(Ag)$ increases rapidly at $\tau\approx4$, almost at the same time instant when the bubble front starts re-accelerating. At later time, $\langle|\omega_0|^2\rangle\lambda/(Ag)$ increases again at $\tau\approx7$ for $Re_p=8000$ and $\tau\approx6$ for $Re_p=20000$, correlating with the onset of a second re-acceleration in the $Fr_B$ plots. Similarly, for $A=0.6$, a positive correlation between the increase in vorticity and the increase in the bubble velocity can also be observed.

The time history of $\omega_0$ within the three disjoint regions inside the bubble is shown in Fig. \ref{fig:omega3Regions}. The vertical distance for each region is $2/(3k)$, as illustrated in Fig. \ref{fig:vor_explain}(b). For the $A=0.04$, $Re_p=20000$ case (left panel in Fig. \ref{fig:omega3Regions}) we can see a similar evolution of vorticity in the three regions but with a time lag. For example, the increase in $\omega_0$ corresponding to the bubble re-acceleration at $\tau\approx4$ first appears in the `far' region at $\tau\approx3.5$, then in the `middle' region at $\tau\approx 4$, and finally in the `near' region at $\tau \approx 4.2$. At $A=0.6$, a similar pattern can also be observed in Fig. \ref{fig:omega3Regions} (right panel).

The strong correlation between vorticity and bubble velocity suggests that  re-acceleration and deceleration of the bubble front is determined by vorticity accumulation inside bubble, consistent with the previous findings \cite{betti2006bubble,ramaprabhu2012late}. Here, we quantitatively show that the vortices which propel the bubble front are not generated inside the bubble, but are generated far below the bubble tip. The vortices then propagate towards the bubble tip. Note that the vortices need to move faster than the bubble tip, which implies that the induced vortical velocity should enhance the advection velocity. 

 Betti and Sanz \cite{betti2006bubble} (see also \cite{ramaprabhu2012late}) proposed a model to predict bubble velocities by taking into account the effects of vorticity 

\begin{gather}
Fr_B = \sqrt{\frac{1}{3\pi} + \frac{r_d}{1-r_d}\frac{|\omega_0|^2}{4\pi kg}} ,
\label{fittedEq}
\end{gather}
where $r_d = \rho_l/\rho_h$. Note that the above model was derived for ablative RTI, where $\omega_0$ is the vorticity transferred by mass ablation from spikes to bubbles \cite{betti2006bubble,yan2016three,Zhangetal18pre}. Ramaprabhu et al. \cite{ramaprabhu2006limits}  applied this model to classical RTI.

Here, we heuristically modify the model by adding an efficiency factor $\eta=0.45$ to the vorticity term to account for the attenuation of vortices as they travel through the bubble tip region. The bubble velocity prediction for 2D RTI is,

\begin{gather}
Fr^{\text{model}}_B = \sqrt{\frac{1}{3\pi} + \eta\frac{r_d}{1-r_d}\frac{|\omega_0|^2}{4\pi kg}}.
\label{fittedEq2}
\end{gather}

\begin{figure*}
\centering
\begin{subfigure}{0.45\textwidth}
\includegraphics[width=2.2 in]{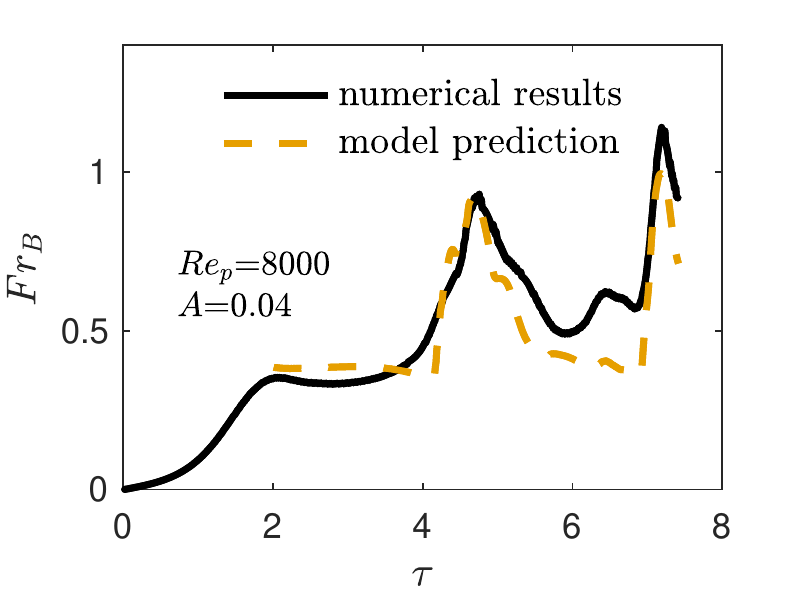}
\end{subfigure}
\begin{subfigure}{0.45\textwidth}
\includegraphics[width=2.2 in]{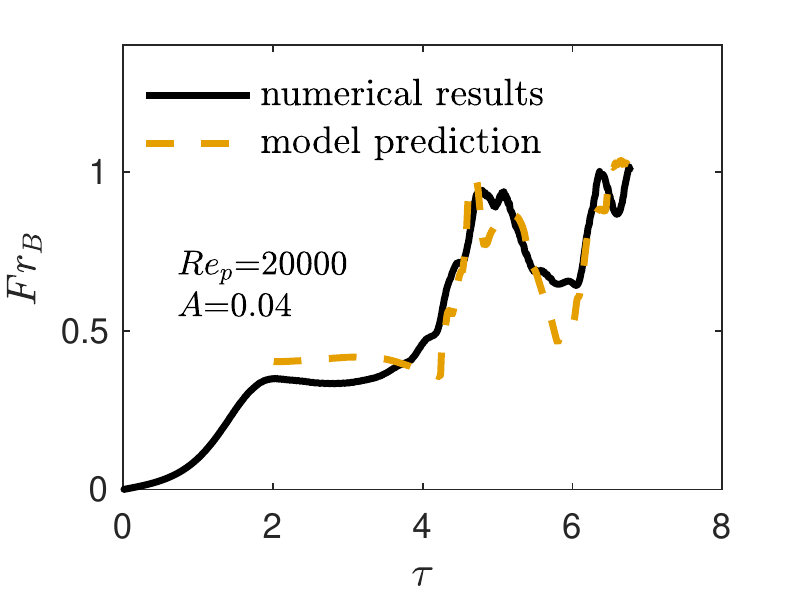}
\end{subfigure}
\\
\vspace{-0.15cm}
\begin{subfigure}{0.45\textwidth}
\includegraphics[width=2.2 in]{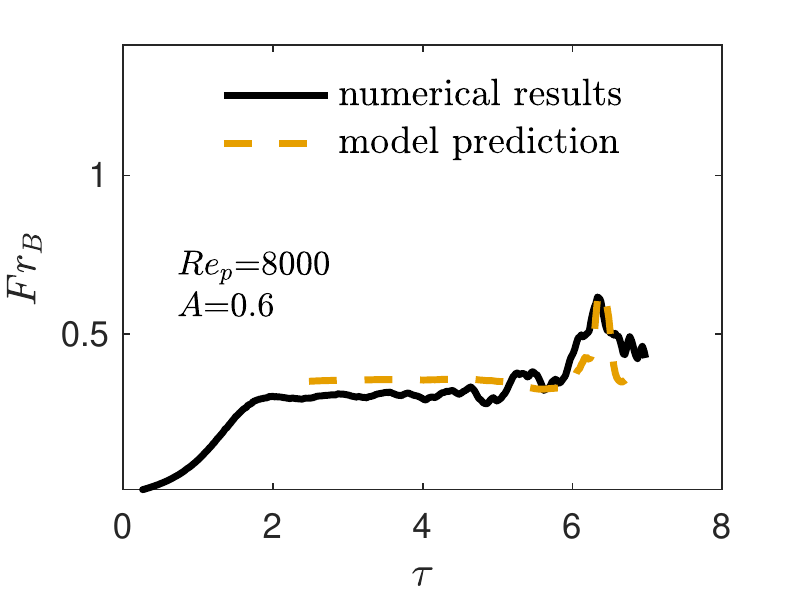}
\end{subfigure}
\begin{subfigure}{0.45\textwidth}
\includegraphics[width=2.2 in]{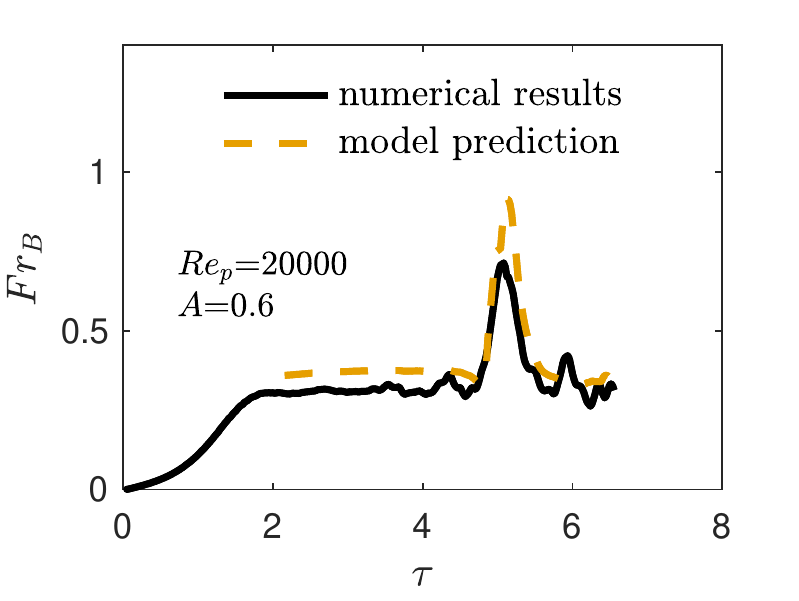}
\end{subfigure}
\renewcommand{\figurename}{FIG.}
\caption{Numerical results of $Fr_B$ and model predictions for four 2D RTI cases, $Re_p=8000$ and $A=0.04$ (top left), $Re_p=20000$ and $A=0.04$ (top right), $Re_p=8000$ and $A=0.6$ (bottom left), $Re_p=20000$ and $A=0.6$ (bottom right). Note that the prediction is only valid in the nonlinear stage. $\eta=0.45$ in all cases. These results corroborate the explanation that vorticity plays a primary role in determining bubble growth in the nonlinear stage.} 
\label{fig:modelPredictVel}
\end{figure*}

Figure  \ref{fig:modelPredictVel} shows that $Fr^{\text{model}}_B$ from  eq. (\ref{fittedEq2}) roughly represents the actual speed $Fr_B$ from the  simulations. Model eq. (\ref{fittedEq2}) captures the bubble velocity development after the potential stage ($\tau \approx 2$) and predicts the onset of bubble re-acceleration. This agreement further supports the explanation that vortices inside the bubble tip are the cause of bubble re-acceleration. 

\section{CONCLUSIONS}
\label{sec:conclusions}
In this paper, we carried out a systematic investigation into the effects of $Re_p$ and $A$ on the development of single-mode RTI in both 2D and 3D. Our simulations used fully compressible dynamics of a single fluid with background temperature variation and uniform  background density. This configuration does not present the instability suppression due to background stratification seen in other studies. The main conclusions are summarized as follows:

1. In 2D RTI, above a threshold $Re_p$ value, the bubble re-accelerates to speeds that are larger than the ``terminal velocity'' predicted in potential flow models. For these high $Re_p$ values, such a speed enhancement is persistent and the bubble does not decelerate at later times as previously suggested \cite{ramaprabhu2012late}. We also observe asymmetric late-time growth in height and speed between bubbles and spikes at low $A$ when $Re_p$ is sufficiently high. 

2. Increasing $A$ while keeping $Re_p$ fixed suppresses the bubble front development. This is in part due to a reduction in secondary instabilities and vortex generation \cite{ramaprabhu2006limits}. However, more importantly, it is because the sites of vortex generation around the sinking spike drift further away from the bubble tip at higher $A$, requiring vortices to travel longer distances before entering the bubble tip region. For a fixed $Re_p$, this leads to the dissipative attenuation of vortices as they travel. The phase diagram in Fig. \ref{fig:exitCond} suggests that if $Re_p$ is sufficiently large, it can counteract the effect of increasing $A$, even in the $A\to1$ limit relevant for ICF. 

3. The effects of $A$ and $Re_p$ on RTI are qualitatively similar in 2D and 3D. However, 3D bubbles are easier to re-accelerate, having a lower $Re_p$ threshold for any $A$. 

4. Analysis of vorticity dynamics shows a clear correlation between vortices inside the bubble tip region and re-acceleration. These vortices are generated around the spike interface as it penetrates into the lighter fluid. We showed how these vortices then propagate towards the bubble tip, eventually causing its re-acceleration. We heuristically modified the Betti-Sanz model \cite{betti2006bubble} for bubble velocity by introducing a vorticity efficiency factor $\eta= 0.45$ to account for the attenuation of vortices as they travel through the bubble tip region.  

We would like to emphasize the significance of maintaining symmetry, which lies in its role as an indicator of momentum conservation. While single-mode RTI rarely occurs in practical applications such as ICF, momentum conservation \emph{does hold} in those more complex systems. In those applications, perturbations are multi-mode and, therefore, momentum conservation is not reflected as the simple symmetry arising in single-modes. Yet, symmetry is important in single-mode RTI because it marks the fidelity of the simulation. If a fundamental conservation law is violated, one must  be (at the very least) skeptical of conclusions drawn from such simulations. This should also raise questions about multi-mode RTI using those codes, because if the simulations violate momentum conservation for single-mode RTI, they are probably also violating it in multi-mode RTI. In this sense, single-mode simulations offer an important benchmark.

We note that further studies are still needed to investigate the late-time behavior of RTI, including higher-resolution simulations capable of reaching higher $Re_p$ and run for longer times to better characterize bubble evolution as $A\rightarrow1$. The latter limit is of particular relevance to ICF, where the extent of bubble growth and penetration into the heavy fluid play a critical role in mixing ablator material into the fuel, with potentially severely adverse effects on the target performance. This study focuses on single-mode RTI, which is a building block for multi-mode RTI. The next step is the study of the effects of $Re_p$ and $A$ on the mixing layer development in multi-mode RTI with practical configurations relevant to ICF.

\section*{Acknowledgements}
This research was funded by LANL LDRD program through project number 20150568ER and DOE FES grants DE-SC0014318 and DE-SC0020229. DZ and HA were also supported DOE NNSA award DE-NA0003856. HA was also supported by NASA grant 80NSSC18K0772 and DOE grant DE-SC0019329.  Computing time was provided by the National Energy Research Scientific Computing Center (NERSC) under Contract No. DE-AC02-05CH11231.

This report was prepared as an account of work sponsored by an agency of the U.S. Government. Neither the U.S. Government nor any agency thereof, nor any of their employees, makes any warranty, express or implied, or assumes any legal liability or responsibility for the accuracy, completeness, or usefulness of any information, apparatus, product, or process disclosed, or represents that its use would not infringe privately owned rights. Reference herein to any specific commercial product, process, or service by trade name, trademark, manufacturer, or otherwise does not necessarily constitute or imply its endorsement, recommendation, or favoring by the U.S. Government or any agency thereof. The views and opinions of authors expressed herein do not necessarily state or reflect those of the U.S. Government or any agency thereof.

\appendix
 \setcounter{figure}{0}
\section{Validation of Simulations with Filtering}
\label{app:filterDNScompare}

The simulations in this work are performed with a sixth-order filter described in \cite{lele1992compact}, which allows us to conduct higher $Re_p$ simulations and for longer times at any given grid resolution. We will now present a comparison with DNS, which rely on the same code but without filtering. The simulations using filtering show very good agreement with DNS. Specifically, results on the bubble velocity, which is the main focus of this study, are almost indistinguishable from those obtained from DNS.

The sixth-order filter \cite{lele1992compact} acts to remove the smallest scales near the grid-scale. Unlike the incompressible RTI formulation \cite{cook2001transition,cabot2006reynolds,livescu2007buoyancy,wei2012late}, the fully compressible dynamics used here introduce much more stringent numerical requirements, which makes DNS very expensive.
For example, to simulate RTI at $Re_p$ = 20000 and $A=0.04$, a DNS that uses a $512 \times 4096$ grid can run until $\tau \approx$ 4.1 before becoming numerically unstable, whereas using a $1024 \times 8192$ grid can run until $\tau \approx$ 4.2. Therefore, the gain from doubling the grid resolution in DNS mode is merely $\Delta\tau\approx0.1$, making the numerical cost prohibitive. A $1024 \times 8192$-grid simulation integrated till the spike reaches the wall consumes $\approx$ 0.3 million CPU hours, implying that conducting DNS for all flows analyzed here is not feasible with the available computing resources.

The main effect of filtering is to regularize the anomalously large $\nabla\cdot \bm u$ values. These appear in a few locations at the interface between heavy and light fluids when RTI develops into the deep nonlinear phase. In our simulations, even though the Mach number $Ma$ is small (e.g. maximum $Ma({\bm x})\approx 0.1$ at $\tau=4$ for the $A=0.04$ $Re_p=8000$ simulation), the acoustic waves generated by the piston-like motions of the bubble and spike can still coalesce into shock waves \cite{olson2007rayleigh}, significantly increasing the dilatation $\nabla\cdot \bm u$. While the role of these shock waves can become important at larger Mach numbers \cite{olson2007rayleigh}, at low Mach numbers, they are essentially decoupled from the rest of the flow, so that filtering them is unlikely to affect the results. To verify this assertion, the filtering results are validated here against DNS at different $A$ and $Re_p$ values. The simulations show no significant differences for both pointwise and spatially averaged  quantities.

\begin{figure}
\centering
\begin{subfigure}{0.05\textwidth}
\includegraphics[width=\textwidth,trim={0cm 0cm 0cm 14cm},clip]{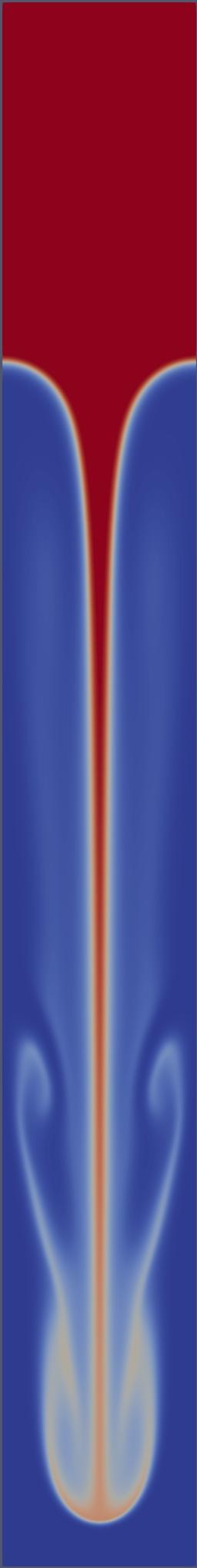}
\caption{}
\end{subfigure}
\begin{subfigure}{0.05\textwidth}
\includegraphics[width=\textwidth,trim={0cm 0cm 0cm 14cm},clip]{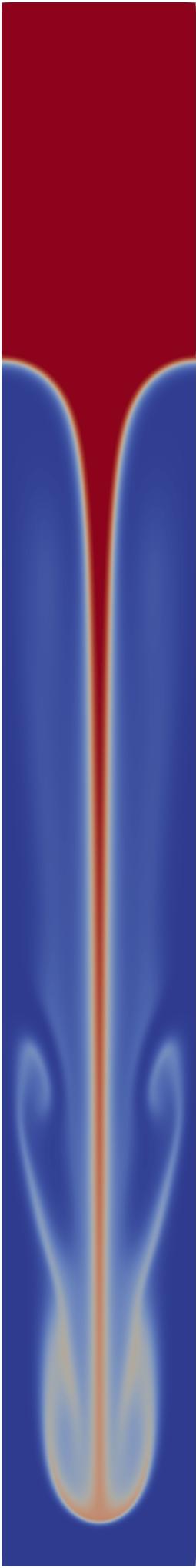}
\caption{}
\end{subfigure}
\begin{subfigure}{0.04\textwidth}
\includegraphics[width=\textwidth]{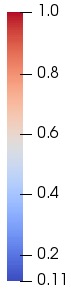}
\caption*{}
\end{subfigure}
\hspace{1cm}
\begin{subfigure}{0.05\textwidth}
\includegraphics[width=\textwidth,trim={0cm 0cm 0cm 14cm},clip]{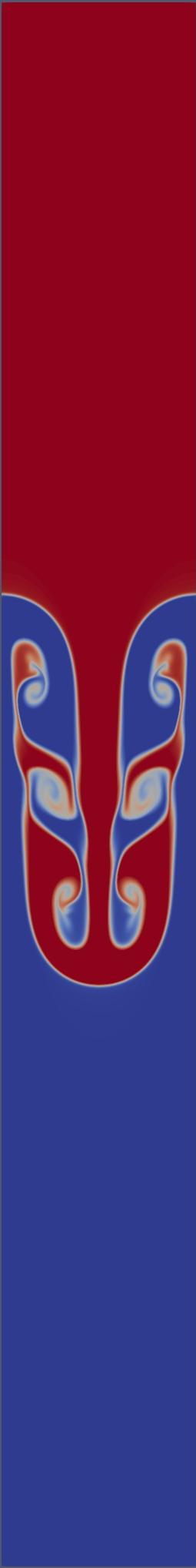}
\caption{}
\end{subfigure}
\begin{subfigure}{0.05\textwidth}
\includegraphics[width=\textwidth,trim={0cm 0cm 0cm 14cm},clip]{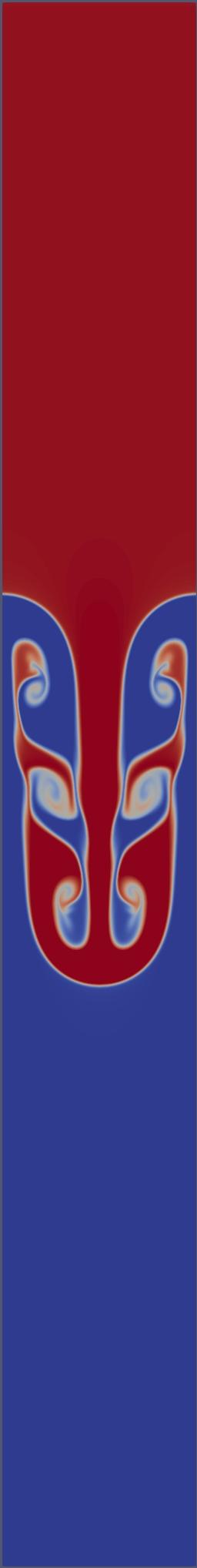}
\caption{}
\end{subfigure}
\begin{subfigure}{0.04\textwidth}
\includegraphics[width=\textwidth]{densityVis/1_083_colorbar.jpg}
\caption*{}
\end{subfigure}
\renewcommand{\figurename}{FIG.}
\caption{Comparison of the density field between DNS and simulations with filtering in 2D RTI. (a) and (b) simulate RTI at $Re_p$=1100 and $A$=0.8 at $\tau$=6. (a) is a DNS on a $256\times2048$ grid, and (b) uses filtering on a $128\times1024$ grid.  (c) and (d) 
simulate RTI at $Re_p$=8000, $A$=0.04 at $\tau$=4. 
(c) is a DNS on a $512\times4096$ grid, and (d) uses filtering on a $256\times2048$ grid. Note that the images are cropped in the vertical to save space (domain aspect ratio is 8).}
\label{fig:filter2Dsingle}
\end{figure}

\begin{figure}
\centering
\begin{subfigure}{0.15\textwidth}
\includegraphics[width=\textwidth]{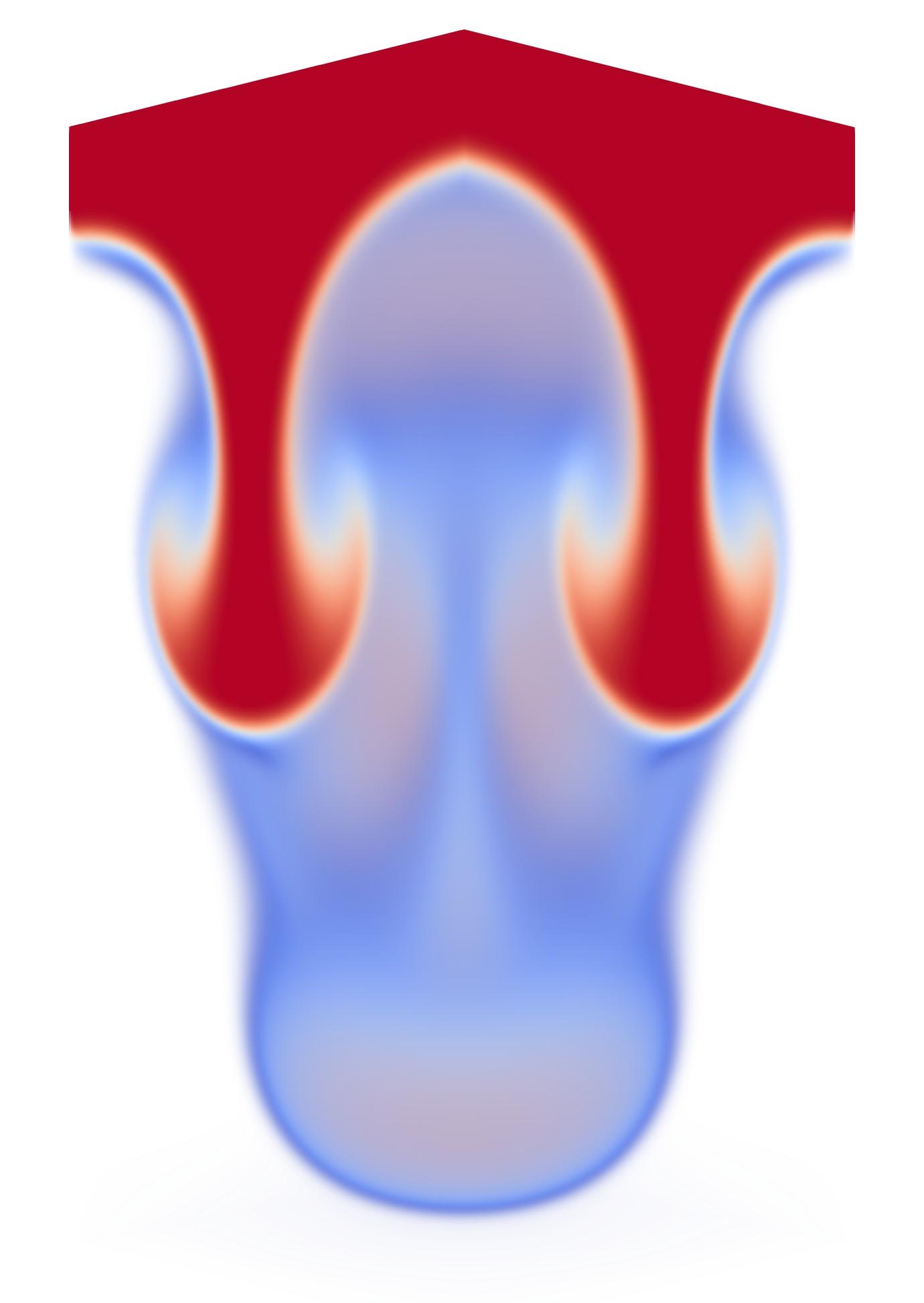}
\caption{DNS}
\end{subfigure}
\begin{subfigure}{0.15\textwidth}
\includegraphics[width=\textwidth]{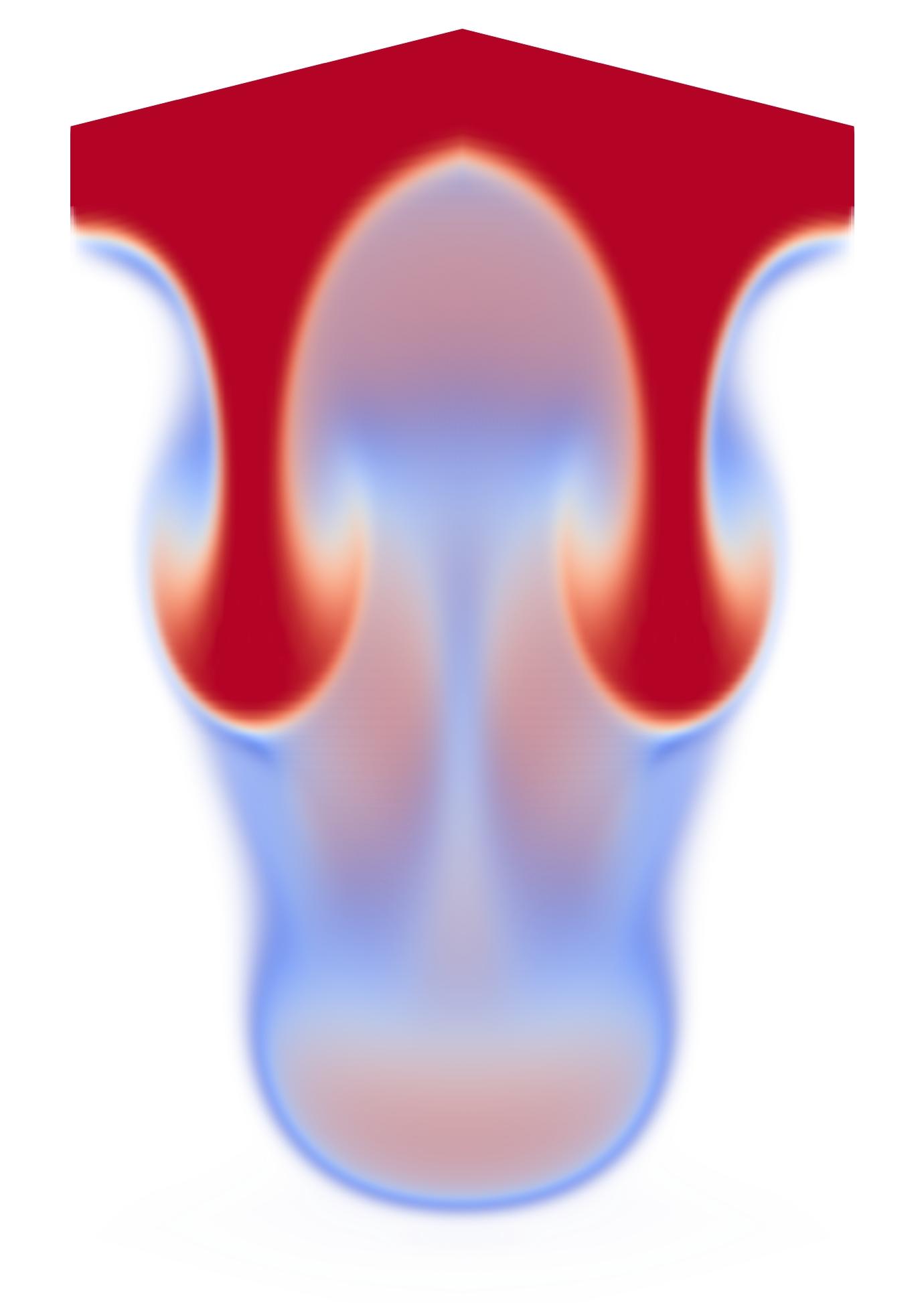}
\caption{filtering}
\end{subfigure}
\begin{subfigure}{0.04\textwidth}
\includegraphics[width=\textwidth]{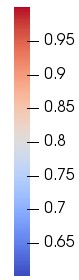}
\caption*{}
\end{subfigure}
\renewcommand{\figurename}{FIG.}
\caption{Comparison of the density field between DNS and simulations with filtering in 3D RTI at $Re_p=800$, $A$=0.25, at $\tau$=2.5. (a) is a DNS on a $128\times128\times256$ grid, and (b) uses filtering on a $64\times64\times128$ grid. The maximum relative difference ($\frac{\text{DNS}-\text{filtering}}{\text{filtering}}$) is 0.29\%. Differences in the color rendering is due to the DNS grid being finer with a higher grid-point density, which leads to slightly darker colors despite the pointwise numerical values being almost identical.}
\label{fig:filter3DSingleA004}
\end{figure}

\begin{figure}
\centering
\begin{subfigure}{0.15\textwidth}
\includegraphics[width=\textwidth]{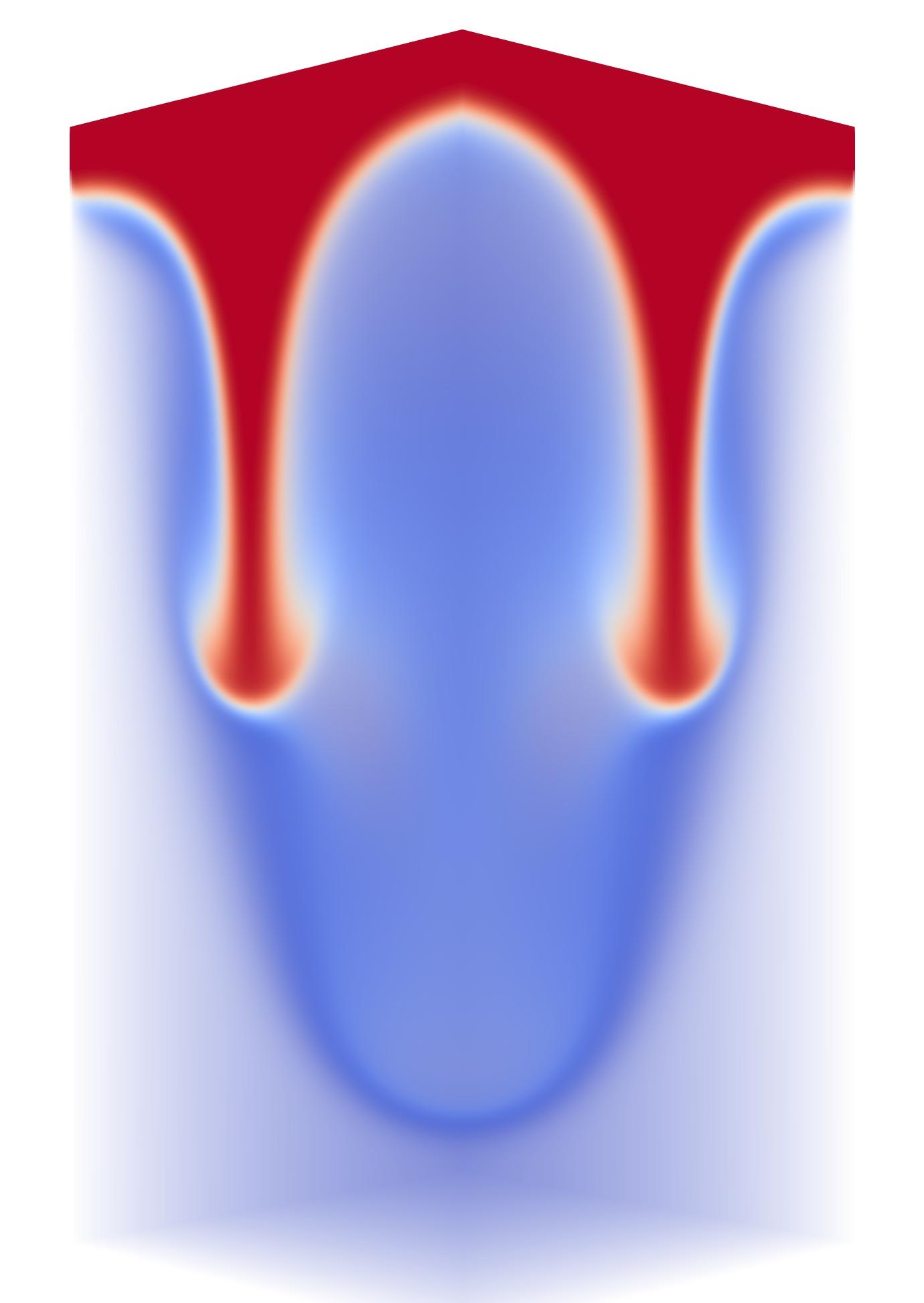}
\caption{DNS}
\end{subfigure}
\begin{subfigure}{0.15\textwidth}
\includegraphics[width=\textwidth]{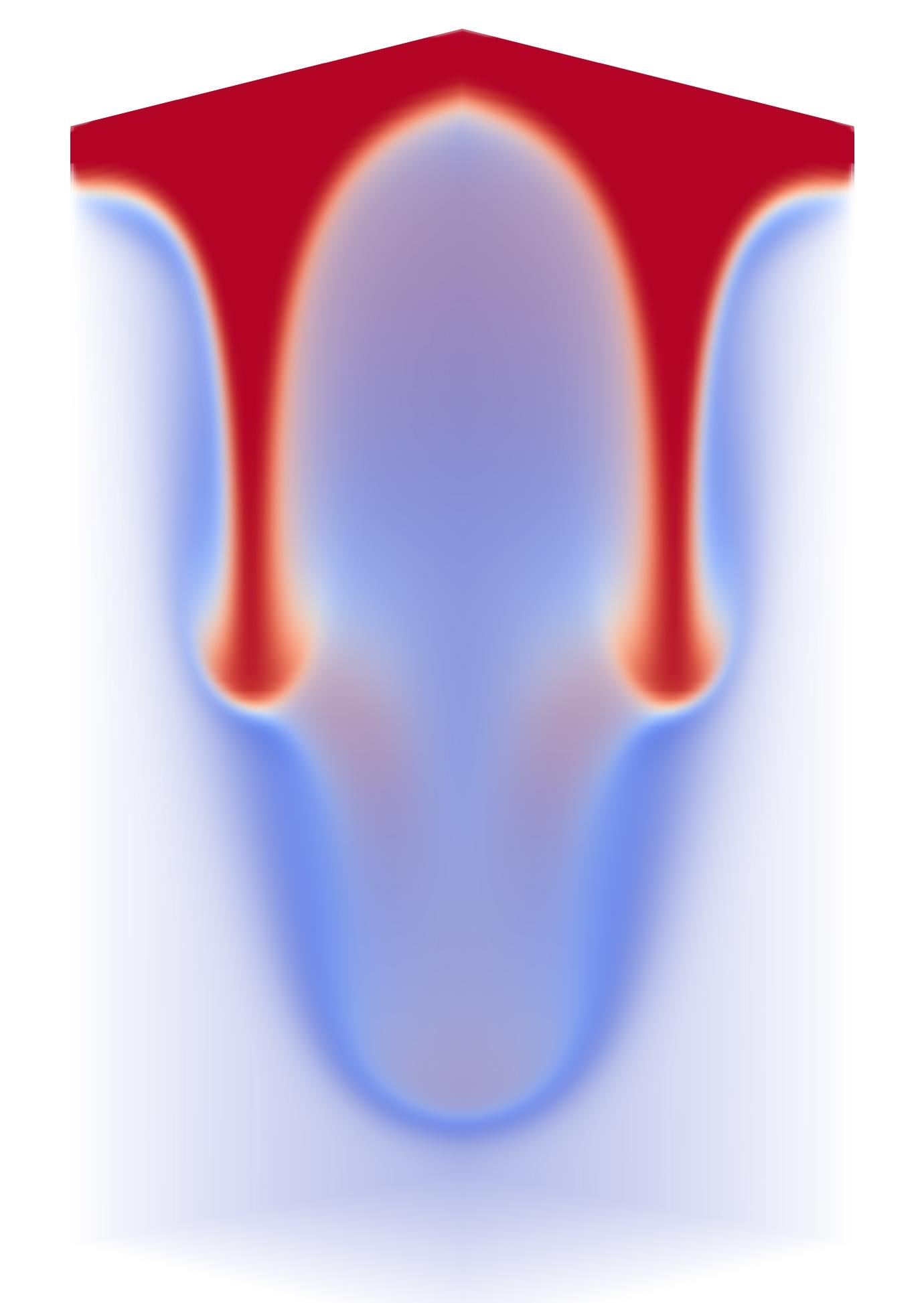}
\caption{filtering}
\end{subfigure}
\begin{subfigure}{0.04\textwidth}
\includegraphics[width=\textwidth]{densityVis/A08_colorbar_appedix.jpg}
\caption*{}
\end{subfigure}
\caption{Comparison of the density field between DNS and simulations with filtering in 3D RTI at $Re_p=450$, $A=0.8$, at $\tau=2.5$. (a) is a DNS on a $128\times128\times256$ grid, and (b) uses filtering on a $64\times64\times128$ grid. The maximum relative difference ($\frac{\text{DNS}-\text{filtering}}{\text{filtering}}$) is 4.78\%. Differences in the color rendering is due to the DNS grid being finer with a higher grid-point density, which leads to slightly darker colors despite the pointwise numerical values being almost identical.}
\label{fig:filter3DSingleA08}
\end{figure}

Figures \ref{fig:filter2Dsingle}-\ref{fig:filter3DSingleA08} compare the density field in both 2D and 3D at different $A$ and $Re_p$. The filtering results are performed at lower grid resolutions than DNS. The plots from DNS and simulations with filtering show almost identical pointwise values. 

\begin{figure}
\caption*{\raggedright\hspace{1.2cm} $\nabla \cdot \bm u$ \hspace{2.8cm} $\omega$ }
\vspace{-0.2cm}
\centering 
\begin{subfigure}{0.055\textwidth}
\includegraphics[width=\textwidth]{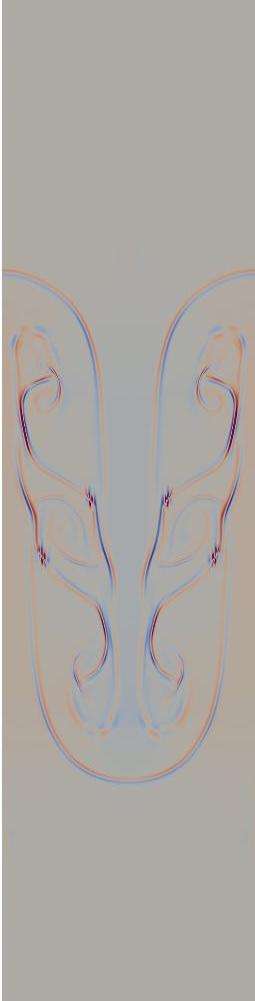}
\caption{}
\end{subfigure}
\begin{subfigure}{0.055\textwidth}
\includegraphics[width=\textwidth]{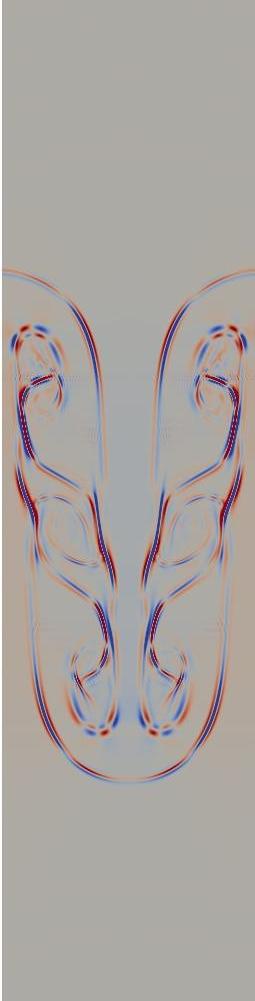}
\caption{}
\end{subfigure}
\begin{subfigure}{0.055\textwidth}
\includegraphics[width=\textwidth]{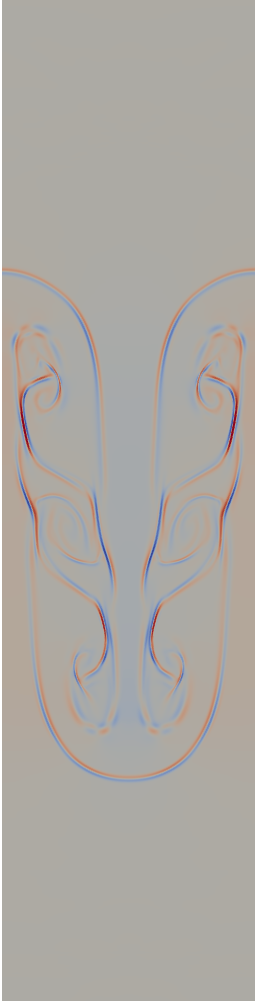}
\caption{}
\end{subfigure}
\begin{subfigure}{0.035\textwidth}
\includegraphics[width=\textwidth]{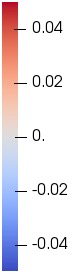}
\caption*{}
\end{subfigure}
\begin{subfigure}{0.055\textwidth}
\includegraphics[width=\textwidth]{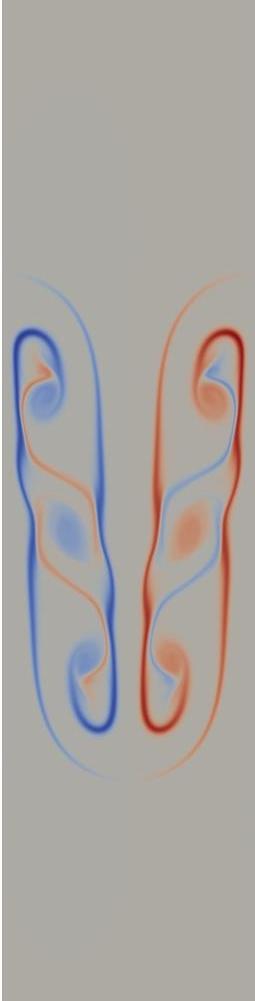}
\caption{}
\end{subfigure}
\begin{subfigure}{0.055\textwidth}
\includegraphics[width=\textwidth]{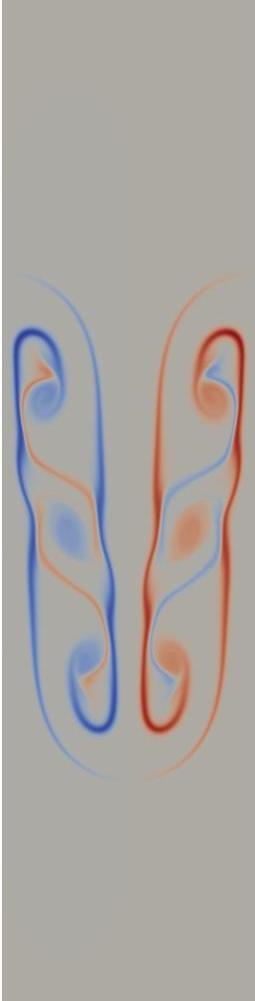}
\caption{}
\end{subfigure}
\begin{subfigure}{0.055\textwidth}
\includegraphics[width=\textwidth]{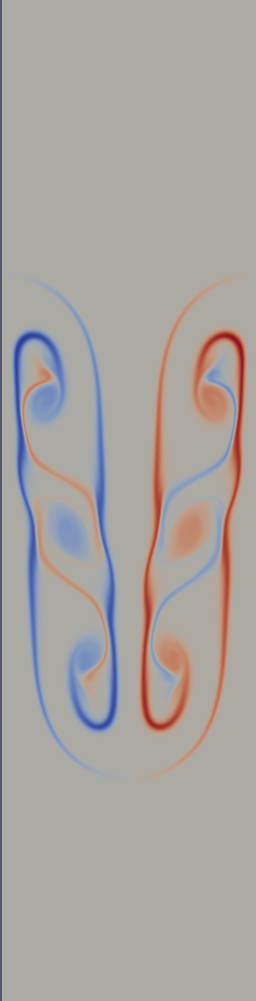}
\caption{}
\end{subfigure}
\begin{subfigure}{0.035\textwidth}
\includegraphics[width=\textwidth]{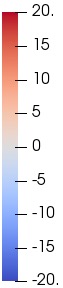}
\caption*{}
\end{subfigure}
\renewcommand{\figurename}{FIG.}
\caption{Comparison of dilatation $\nabla \cdot \bm u$ and vorticity $\omega$ between DNS and simulations with filtering in 2D RTI at $Re_p=8000$, $A=0.04$, at $\tau=4$. (a, d) are DNS on a $512\times4096$-grid. (b, e) use filtering on a coarser $256\times2048$-grid. (c, f) use filtering on a $512\times4096$-grid. Images are cropped vertically to save space (domain aspect ratio is 8).}
\label{ptcomp}
\end{figure}

Figure \ref{ptcomp} shows a pointwise comparison of $\nabla\cdot \bm u$ and $\omega$. When using the same grid resolution ($512\times4096$), DNS and simulations with filtering yield 
almost identical $\nabla\cdot \bm u$ and $\omega$ fields. However, a close inspection of Figs. \ref{ptcomp}(a),(c) shows how filtering reduces $\nabla\cdot \bm u$ when compared to DNS. Lower-resolution ($256\times2048$) simulations with filtering still yield $\omega$ that is indistinguishable from that of DNS, but $\nabla\cdot \bm u$ is noticeably larger. We attribute the larger $\nabla\cdot \bm u$ values on a coarser grid (but the same $Re_p$) to an insufficient resolution to accurately evolve the dilatation field. Without filtering, $\nabla\cdot \bm u$ can become very large in an under-resolved simulation leading to numerical instabilities. Filtering keeps the values of $\nabla\cdot \bm u$ bounded, without affecting other aspects of the RTI dynamics. Again, this is consistent with the low Mach numbers regimes attained in the simulations, for which the shocklets \cite{olson2007rayleigh} become decoupled from the rest of the flow. More detailed discussions of filtering can be found in \cite{lele1992compact}.

\begin{figure*}[!h]
\centering
\begin{subfigure}{0.32\textwidth}
\includegraphics[width=\textwidth]{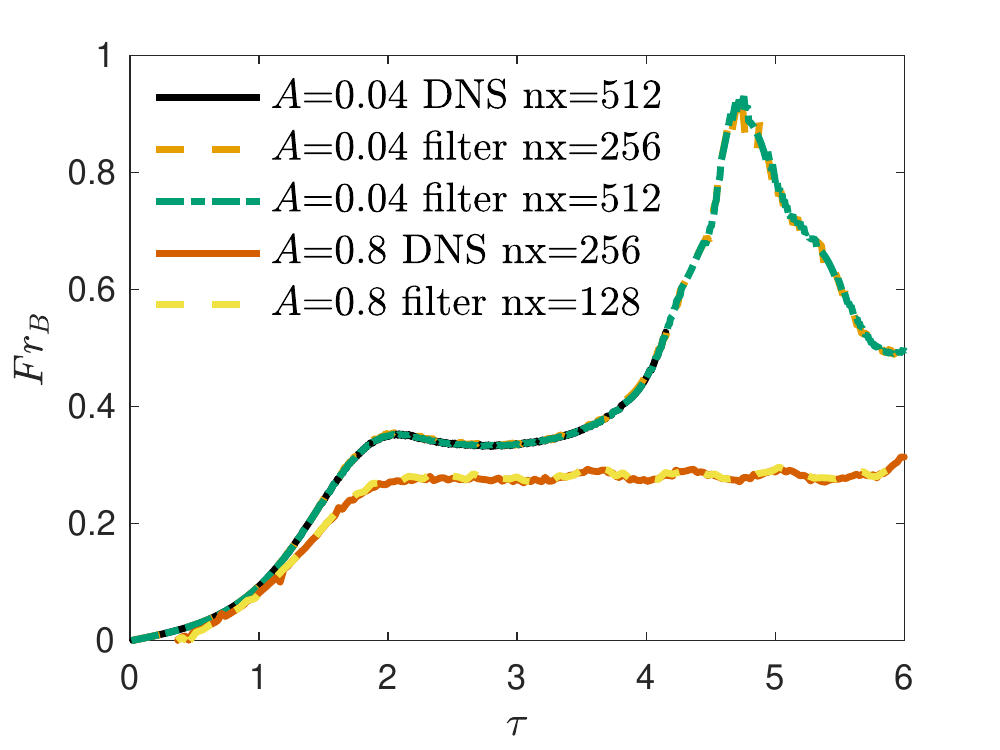}
\end{subfigure}
\begin{subfigure}{0.32\textwidth}
\includegraphics[width=\textwidth]{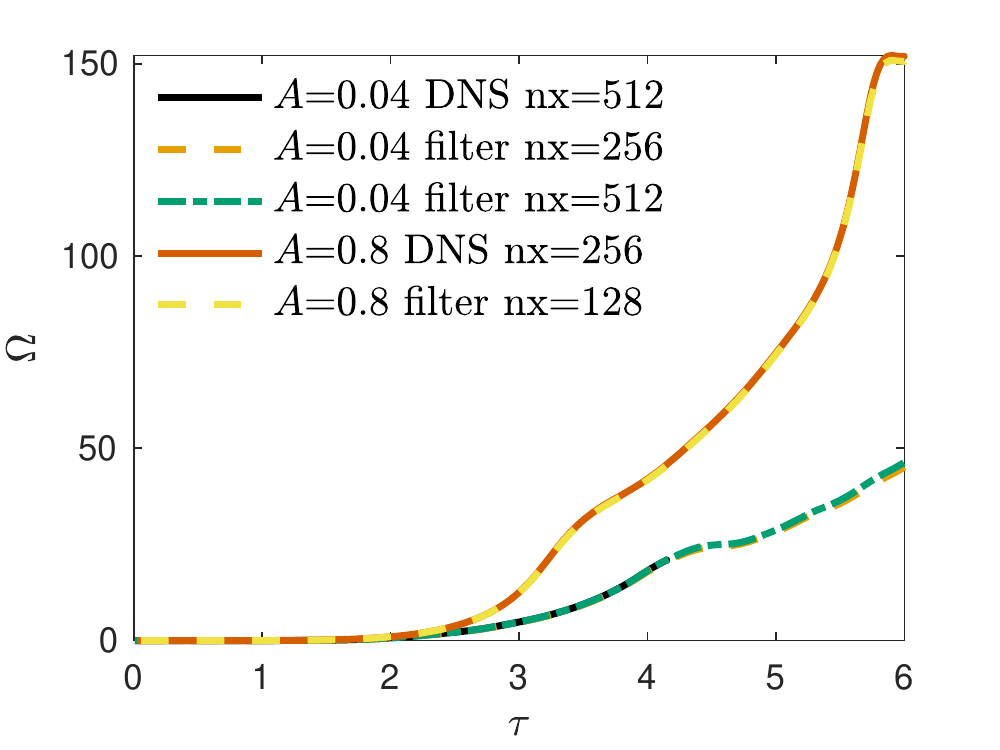}
\end{subfigure}
\\
\begin{subfigure}{0.32\textwidth}
\includegraphics[width=\textwidth]{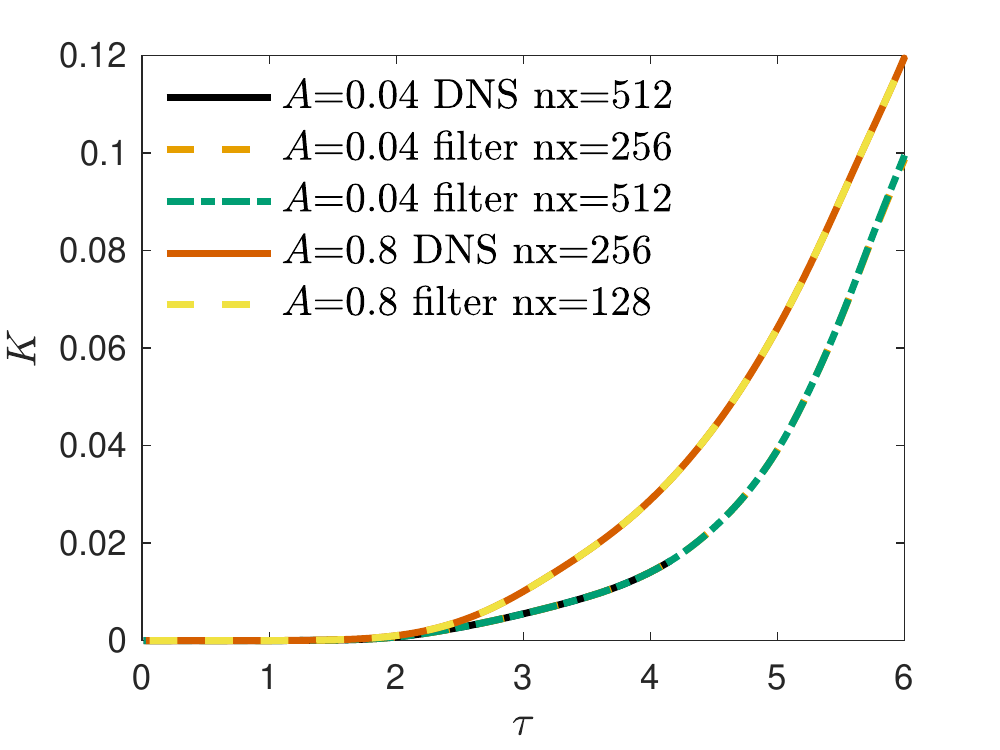}
\end{subfigure}
\begin{subfigure}{0.32\textwidth}
\includegraphics[width=\textwidth]{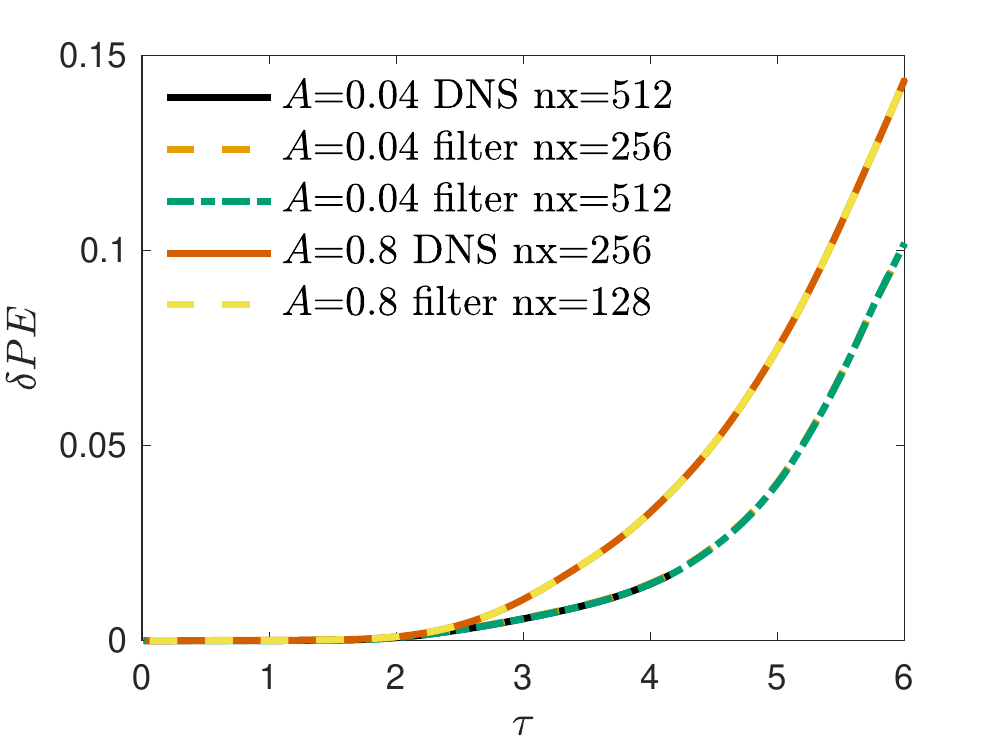}
\end{subfigure}
\begin{subfigure}{0.32\textwidth}
\includegraphics[width=\textwidth]{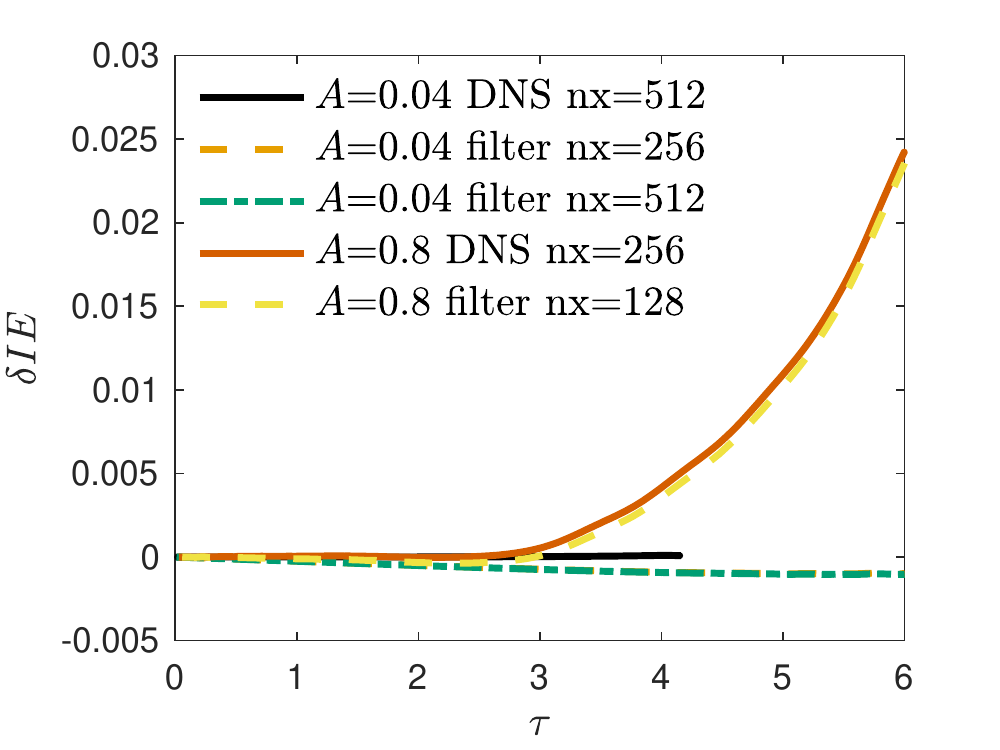}
\end{subfigure}
\renewcommand{\figurename}{FIG.}
\caption{
Comparison of bubble velocity $Fr_B$ (top left), enstrophy $\Omega$ (top right), kinetic energy $K$ (bottom left), release of potential energy $\delta PE$ (bottom middle), and change in internal energy $\delta IE$ (bottom right) between DNS and filtering results for 2D RTI. 
RTI is simulated at $A=0.04$, $Re_p=8000$ using DNS on a 512$\times$4096 grid, and filtering on 256$\times$2048  and 512$\times$4096 grids.
We also show RTI at $A=0.8$, $Re_p=1100$ using DNS on a 256$\times$2048-grid and filtering on a 128$\times$1024-grid.
Note that the plots of $K$, $\delta P$, and $\Omega$ for $A=0.04$ are multiplied by 5.}
\label{filter2DSingleQuan}
\end{figure*}

In addition to the pointwise comparison of visualizations, a comparison of several quantities as a function of time is presented in Figure  \ref{filter2DSingleQuan}. These are the bubble velocity $Fr_B(\tau)$, kinetic energy $K(\tau) = \int_V\rho |\bm u|^2/2\,dV$, released potential energy $\delta PE(\tau)=\int_V [\rho(\bm x, 0) - \rho(\bm x, \tau)]\, g\, z\, dV$, enstrophy $\Omega(\tau) = \int_V |\bm \omega|^2 dV$, and change in internal energy $\delta IE(\tau)=\int_V [\rho e(\bm x, \tau) - \rho e(\bm x, 0)]\, dV$. The comparison between DNS and filtered simulations shows an almost identical evolution for four out of the five quantities. Small differences are discernible in the evolution of $\delta IE$, but even these are miniscule relative to the magnitude. The small differences indicate that filtering yields slightly smaller $\delta IE$ compared to DNS due to removal of the energy dissipated by shocklets. 

In summary, simulations with filtering seem to yield results that are almost identical to DNS at higher resolutions. This is especially the case for the bubble velocity, the focus of the present paper.

\section{Simulation Parameters}
\label{app:simpara}
All simulation parameters in the main document are summarized in Tables \ref{tbl:2dSimpara} and \ref{tbl:3dSimpara}. 

\begin{table}[!h]
\centering
\caption{2D simulation parameters. I.Loc. means the initial interface location. Runs 1-3 are DNS. All others use filtering (see  \ref{app:filterDNScompare}).}
\label{tbl:2dSimpara}
\begin{tabular}{lllllll}
\hline
\hline
   Run         & $Re_p$   & Grid Size   &$Gr_{\Delta}$    & $A$           & I.Loc. \\ \hline
   $\rm 1$   & 100 &   $128\times1024$  & 0.0099    & 0.04        & $L_z/2$ \\
   $\rm 2$   & 210 &   $128\times1024$  & 0.044    & 0.04         & $L_z/2$  \\
   $\rm 3$   & 400 &   $128\times1024$  & 0.16    & 0.04          & $L_z/2$   \\
   $\rm 4$  & 1000 &  $128\times1024$   & 0.99    & 0.04          & $L_z/2$ \\
   $\rm 5$  & 1500 &  $256\times2048$   & 0.28    & 0.04          & $L_z/2$   \\ 
   $\rm 6$  & 2000 &  $256\times2048$   & 0.50     & 0.04          & $L_z/2$ \\
   $\rm 7$  & 4000 &  $512\times4096$   & 0.25      & 0.04          & $L_z/2$ \\
   $\rm 8$  & 6000&   $512\times4096$   & 0.56    & 0.04             & $L_z/2$ \\
   $\rm 9$  & 8000&   $512\times4096$   & 0.99    & 0.04             & $L_z/2$ \\
   $\rm 10$ & 20000&  $1024\times8192$  & 0.77    & 0.04             & $L_z/2$  \\
   $\rm 11$  & 1000 &  $128\times1024$  & 1.19   & 0.25          & $L_z/2$ \\
   $\rm 12$  & 1500 &  $256\times2048$  & 0.33    & 0.25          & $L_z/2$   \\ 
   $\rm 13$  & 2000 &  $256\times2048$  & 0.60   & 0.25          & $L_z/2$ \\
   $\rm 14$  & 4000 &  $512\times4096$  & 0.30   & 0.25          & $L_z/2$ \\
   $\rm 15$ & 6000&   $512\times4096$   & 0.67   & 0.25              & $L_z/2$    \\
   $\rm 16$ & 8000&   $512\times4096$   & 1.19   & 0.25              & $L_z/2$    \\
   $\rm 17$ & 15000&   $512\times4096$  & 4.19   & 0.25              & $L_z/2$    \\
   $\rm 18$ & 20000&  $512\times4096$   & 7.45   & 0.25              & $L_z/2$    \\
   $\rm 19$ & 4000&   $512\times4096$   & 0.32  & 0.35              & $5L_z/8$   \\
   $\rm 20$ & 8000&   $512\times4096$   & 1.29   & 0.35              & $5L_z/8$   \\
   $\rm 21$ & 10000&   $512\times4096$  & 2.01   & 0.35              & $5L_z/8$    \\
   $\rm 22$ & 15000&   $512\times4096$  & 4.53   & 0.35              & $5L_z/8$   \\
   $\rm 23$ & 20000&   $512\times4096$  & 8.05   & 0.35              & $5L_z/8$    \\
   $\rm 24$ & 4000&   $512\times4096$   & 0.35  & 0.45              &  $5L_z/8$ \\
   $\rm 25$ & 6000&   $512\times4096$   & 0.78   & 0.45              & $5L_z/8$    \\
   $\rm 26$ & 8000&   $512\times4096$   & 1.38   & 0.45              &  $5L_z/8$    \\
   $\rm 27$ & 15000&   $512\times4096$  & 4.86  & 0.45              &  $5L_z/8$    \\
   $\rm 28$ & 20000&   $512\times4096$  & 8.64  & 0.45              &  $5L_z/8$    \\
   $\rm 29$ & 20000&   $512\times4096$  & 9.06  & 0.52              &$3L_z/4$    \\
   $\rm 30$ & 2000&   $256\times2048$   & 0.76  & 0.6              & $3L_z/4$    \\
   $\rm 31$ & 4000&   $512\times4096$   & 0.38   & 0.6              & $3L_z/4$    \\
   $\rm 32$ & 6000&   $512\times4096$   & 0.85   & 0.6              & $3L_z/4$    \\
   $\rm 33$  & 8000 &  $512\times4096$  & 1.52     & 0.6             & $3L_z/4$ \\ 
   $\rm 34$  & 20000 & $512\times4096$  & 9.54     & 0.6             & $3L_z/4$   \\
   $\rm 35$ & 2000&   $256\times2048$   & 0.86  & 0.8               & $3L_z/4$    \\
   $\rm 36$ & 4000&   $512\times4096$   & 0.43  & 0.8               & $3L_z/4$    \\
   $\rm 37$ & 6000&   $512\times4096$   & 0.97   & 0.8               & $3L_z/4$    \\
   $\rm 38$  & 8000 &  $512\times4096$  & 1.71     & 0.8              & $3L_z/4$  \\ 
   $\rm 39$  & 20000 & $512\times4096$  & 10.73   & 0.8             & $3L_z/4$   \\ 
   $\rm 40$  & 30000 & $1024\times8192$ & 3.02   & 0.8             & $3L_z/4$   \\ 
    \hline  
\end{tabular}
\end{table}

\begin{table}[!h]
\centering
\caption{3D simulation parameters. I.Loc. means the initial interface location. Runs 41-43 are DNS. All others use filtering (see  \ref{app:filterDNScompare}).}
\label{tbl:3dSimpara}
\begin{tabular}{lllllll}
\hline
\hline
   Run         & $Re_p$   & Grid Size   &$Gr_{\Delta}$    & $A$     & I.Loc.        \\ \hline
   $\rm 41$  & 100 &    $128\times128\times1024$    & 0.0099     & 0.04       & $L_z/2$   \\
   $\rm 42$  & 400 &   $128\times128\times1024$    & 0.16     & 0.04          & $L_z/2$   \\
   $\rm 43$  & 1000 &  $128\times128\times1024$    & 0.99     & 0.04         & $L_z/2$   \\ 
   $\rm 44$  & 8000 &  $256\times256\times2048$  &  7.93   & 0.04                & $L_z/2$   \\ 
   $\rm 45$  & 1000 &  $128\times128\times1024$  & 1.19     & 0.25            & $L_z/2$   \\ 
   $\rm 46$  & 1000 &  $128\times128\times1024$  &  1.53   & 0.6                & $3L_z/4$   \\ 
   $\rm 47$  & 1000 &  $128\times128\times1024$  &   1.72   & 0.8             & $3L_z/4$   \\ 
   $\rm 48$  & 8000 &  $256\times256\times2048$   &   13.73   & 0.8            & $3L_z/4$   \\ 
    \hline  
\end{tabular}
\end{table}

\clearpage
\bibliographystyle{unsrt}
\bibliography{singleRT.bib}

\end{document}